\documentclass[]{jfm}

\usepackage{graphicx}
\usepackage{newtxtext}
\usepackage{newtxmath}
\usepackage{natbib}
\usepackage{hyperref}
\usepackage{url}
\usepackage{float}
\usepackage{graphicx}
\usepackage{epstopdf, epsfig}
\usepackage{amsmath}
\usepackage{empheq}
\usepackage{multicol}
\usepackage{subfig}
\usepackage{bm}
\usepackage{accents}
\usepackage{bigints}
\usepackage{booktabs}
\usepackage{array}
\usepackage[most]{tcolorbox}
\usepackage{amsmath}

\hypersetup{
    colorlinks = true,
    urlcolor   = blue,
    citecolor  = black,
}

\newcommand{\RomanNumeralCaps}[1]
\newcolumntype{P}[1]{>{\centering\arraybackslash}p{#1}}

\newtcbox{\mymath}[1][]{%
	nobeforeafter, math upper, tcbox raise base,
	enhanced, colframe=blue!30!black,
	colback=blue!30, boxrule=1pt,
	#1}

\shorttitle{Notes}
\shortauthor{Patankar and Basak and Dasgupta}

\title[Dynamic stabilisation of RP modes]{Dynamic stabilisation of Rayleigh-Plateau modes on a liquid cylinder}

\author{Sagar Patankar\aff{1},
	Saswata Basak\aff{1},
	\and Ratul Dasgupta\aff{1}\corresp{\email{dasgupta.ratul@iitb.ac.in}}}

\affiliation{\aff{1}Department of Chemical Engineering, Indian Institute of Technology, Bombay, Powai 400076, India}

\begin{document}
\maketitle
	\newcommand{\mk}{{\mathup{K}}}
\newcommand{\mi}{{\mathrm{I}}}

	\newcommand*{\dt}[1]{%
	\accentset{\mbox{\large\bfseries .}}{#1}}
\newcommand{\Ou}{^{\mathcal{O}}}
\newcommand*{\ddt}[1]{%
	\accentset{\mbox{\large\bfseries ..}}{#1}}

\begin{abstract}
	We demonstrate dynamic stabilisation of axisymmetric Fourier modes susceptible to the classical Rayleigh-Plateau (RP) instability on a liquid cylinder by subjecting it to a radial oscillatory body force. Viscosity is found to play a crucial role in this stabilisation. Linear stability predictions are obtained via Floquet analysis demonstrating that RP unstable modes can be stabilised using radial forcing. We also solve the linearised, viscous initial-value problem for free-surface deformation obtaining an equation governing the amplitude of a three-dimensional Fourier mode. This equation generalises the Mathieu equation governing Faraday waves on a cylinder derived earlier in \cite{patankar2018faraday}, is non-local in time and represents the cylindrical analogue of its Cartesian counterpart \citep{beyer1995faraday}. The memory term in this equation is physically interpreted and it is shown that for highly viscous fluids, its contribution can be sizeable. Predictions from the numerical solution to this equation demonstrates RP mode stabilisation upto several hundred forcing cycles and is in excellent agreement with numerical simulations of the incompressible, Navier-Stokes equations.
\end{abstract}

{\bf MSC Codes }  {\it(Optional)} Please enter your MSC Codes here

	\section{Introduction} 
Liquid cylinders, jets or annular liquid films coating rods often deform or fragment into a series of droplets of unequal sizes via the ubiquitous Rayleigh-Plateau (RP hereafter) capillary mechanism \citep{plateau1873statique,rayleigh1892xvi}. This may easily be seen, for example, in a jet issuing out of a faucet \citep{rutland1971non}, in a capillary liquid bridge held between two disks \citep{plateau1873statique} or in a film coating a rod \citep{goren1962instability}, to mention but a few situations. Depending on the application, droplet formation may be desirable or it might even be necessary to suppress it. When breakup is intended (e.g. in microfluidic devices cf. \cite{stone2004engineering} or drop-on-demand inkjet printing cf. \cite{driessen2013drop}), strategies are sought such that the size distribution of the resultant droplets and their spacing are controllable e.g. \cite{driessen2014control}. Conversely, when breakup is undesirable stabilisation strategies are necessary and a number of techniques have been proposed towards this. Table \ref{tab_lit} provides a broad summary of known techniques of RP stabilisation and it is apparent that this continues to be an active area of current research. 

The purpose of the present study is to demonstrate dynamic stabilisation of unstable RP modes on a liquid cylinder by subjecting the cylinder to a radial, sinusoidal-in-time body force. It is demonstrated analytically that this is possible and that viscosity plays a crucial role in this stabilisation. The \textit{viscous} analysis presented here significantly builds upon the inviscid analysis presented earlier in \cite{patankar2018faraday} where dynamic stabilisation of RP modes was also predicted but was found to be extremely short-lived in inviscid simulations. In contrast to our earlier inviscid study \citep{patankar2018faraday}, we demonstrate here that for a viscous liquid, by carefully tuning the strength and frequency of (radial) forcing, RP modes accessible to the system maybe rendered stable thus stabilising the cylinder for long time (many forcing time periods). The theoretically predicted stabilisation is verified using numerical simulations of the Navier-Stokes equations demonstrating excellent agreement. 

The study is organised as follows: in subsection \ref{sec_1.1} a brief literature survey discussing the gamut of stabilisation strategies for finite and infinitely long liquid cylinders alongwith a brief background of parametric instabilities and dynamic stabilisation strategies is presented. In section $2$, linear stability analysis of an infinite cylinder of viscous liquid subject to a radial, oscillatory body force is reported via Floquet analysis. Section $3$ reports the derivation of a novel integro-differential equation governing the linearised amplitude of surface modes. The theoretically predicted stabilisation in section \ref{sec_stab} is verified using numerical simulations of the incompressible Navier-Stokes equations (DNS) in section \ref{sec_dns}. The integro-differential equation is physical interpreted and the significance of the memory term are discussed are discussed at the end of section \ref{sec_dns}. Conclusions are discussed in section $6$.

	\subsection{Literature review}\label{sec_1.1}
Stabilisation of RP modes for liquid cylinders are typically investigated either in the context of bridges of finite length or in the infinitely long cylinder approximation. We recall that a cylindrical liquid bridge of length $L$ and diameter $d$ in neutrally buoyant surroundings is stable for slenderness ratio $L/d \leq \pi$ also known as the Plateau limit, see \cite{plateau1873experimental}. Electric field has long been used to both generate stable cylindrical jets \citep{taylor1969electrically} and to stabilize liquid bridges composed of dielectric fluids \citep{raco1968electrically,sankaran1993experiments,thiessen2002active}. Alternatively, application of axial magnetic fields \citep{nicolas1992magnetohydrodynamic} or flow induced stabilisation techniques \citep{lowry1997stability,lowry1994stabilization,lowry1995flow} have been utilised for surmounting the Plateau limit, obtaining stabilisation upto $L/R=8.99$ for a pinned liquid bridge. Another class of techniques comprise acoustic forcing which have been used to demonstrate stabilisation of liquid bridges beyond the Plateau limit \citep{marr1997stabilization,marr2001passive}. The nonlinear dynamics of liquid bridges and their stability subject to axial, oscillatory forcing of the point of support have in fact been studied quite extensively \citep{chen1993nonlinear,mollot1993nonlinear,benilov2016stability,haynes2018stabilization}. Analogously, the use of axial vibration for stabilising and preventing rupture of a thin film coating a solid rod by subjecting one end of the rod to ultrasound forcing has been investigated in detail \citep{ moldavsky2007dynamics,rohlfs2014stabilizing,binz2014direct}. Parametric stabilisation also known as dynamic stabilisation via imposition of vibration has been demonstrated \citep{wolf1970dynamic} for the Rayleigh-Taylor instability of a heavier fluid overlying a lighter one. Here viscosity was found to be crucial for stabilisation of short wavelength modes. In this study we will find that an identical situation occurs in the dynamic stabilisation of RP modes also. Here short wavelength modes (i.e those with wavelength smaller than the cylinder circumference) which are stable in the absence of forcing can however become unstable in the presence of forcing. These modes even when absent in the initial conditions can be produced due to nonlinearity (in numerical simulations) and it will be seen that viscosity is crucial in preventing destabilisation of the cylinder due to these modes.

Parametric stabilisation and destabilisation of otherwise unstable or stable mechanical equilibria have a long and distinguished history of investigation. The first problems to be investigated were mechanical systems, notably by \cite{melde1860ueber} who studied transverse oscillations of a taut string whose end was subjected to lengthwise vibrations (see  \cite{tyndall1901sound}, section 7, figs. 45-49). In a series of studies \cite{rayleigh1883xxxiii,rayleigh1887xvii}, \cite{matthiessen1868akustische} and  \cite{raman1909maintenance,raman1912experimental} studied this problem in detail obtaining the damped Mathieu equation already in their analysis. Closely related experimental observations for fluid interfaces (using mercury, egg-white, turpentine oil etc.) had been made nearly thirty years earlier by \cite{faraday1837peculiar} culminating in the insightful study by \cite{benjamin1954stability} of the instability, which in modern parlance  has come to be known as the Faraday instability.

\cite{benjamin1954stability} derived the Mathieu equation from the inviscid, irrotational fluid equations opening the way to a rich body of literature on Faraday waves \citep{kumar1994parametric,cerda1997faraday,fauve1998waves,PhysRevE.62.1416,adou2016faraday}, spatio-temporal chaos \citep{kudrolli1996patterns}, wave turbulence \citep{shats2014turbulence,holt1996faraday} and pattern-formation \citep{edwards1994patterns,arbell2000temporally}. Viscosity constitutes a non-trivial modification to the Mathieu equation. Unlike inviscid predictions on the forcing-strength versus wavenumber plane, the threshold acceleration for the instability becomes finite when viscosity is taken into account, as the instability tongues do not touch the wavenumber axis anymore. This was first systematically demonstrated by \cite{kumar1994parametric} using Floquet analysis further finding that the wavelength at the onset of the instability varies non-monotonically with increasing viscosity. The predictions of \cite{kumar1994parametric} have been validated in experiments by \cite{bechhoefer_ego_manneville_johnson_1995} and for Faraday waves in a cylinder by \cite{batson_zoueshtiagh_narayanan_2013}. 

The stability tongues of the Mathieu equation suggest the possibility of dynamical stabilisation of a statically unstable configuration of heavier fluid on a top of a lighter one via high-frequency oscillation normal to the unperturbed interface. Since the theoretical and experimental demonstration of this by \cite{wolf1969dynamic,wolf1970dynamic}, this has been studied extensively not only for the Rayleigh-Taylor instability \citep{troyon1971theory,piriz2010dynamic,boffetta2019suppression} but also in the suppresion of long surface-gravity modes in inclined plane flow \citep{woods1995instability}, the Marangoni instability \citep{thiele2006long} and for stabilising a thin film on the underside of a substrate \citep{sterman2017rayleigh}. In close analogy to the work of \cite{wolf1970dynamic}, our present study demonstrates usage of radial forcing (i.e. normal to the unperturbed interface) for dynamic stabilisation of RP modes. To the best of our knowledge, this is the first such demonstration (a condensed version was presented in \cite{patankar2019fragmenting} and \cite{GFM19}). We closely follow the Floquet analysis approach of \cite{kumar1994parametric} in order to obtain the threshold forcing where RP mode stabilisation can be achieved. For viscous liquid cylinders, a recent study by \cite{maity2021floquet} has investigated via Floquet analysis, the effect of viscosity on the stability tongues of the inviscid Mathieu equation proposed in \cite{patankar2018faraday} and investigated further in \cite{maity2020instability}. An interesting observation here is that the $m=1$ mode shows a threshold which decreases with increasing viscosity, in a certain window of viscosity change \citep{maity2021floquet}. The study by \cite{maity2021floquet} however did not investigate the possibility of stabilisation of RP unstable modes, as is the focus of the current study. 

For Faraday waves on flat interfaces, prior studies have demonstrated that the viscous extension of the inviscid Mathieu equation \citep{benjamin1954stability} is an integro-differential equation \citep{jacqmin1988instabilities, beyer1995faraday,cerda1997faraday,cerda1998faraday}. In this study, we also derive a novel cylindrical analogue of this integro-differential equation governing small-amplitude Fourier modes on a liquid cylinder and demonstrate its connection to the equation derived earlier by \cite{beyer1995faraday}. Numerical solution to this integro-differential equation enables us to estimate the contribution of viscosity from the potential part of the flow and from the boundary layer at the free-surface. Additionally, the solution to this equation demonstrates the RP stabilisation that is sought, in excellent agreement with direct numerical simulations. 
\begin{table}
	\begin{center}
		\begin{tabular}{lclc}
			\textbf{Stabilisation technique}  & \textbf{References}  & \textbf{Comments} & \\[3pt]
			Electric field  & \cite{raco1968electrically,sankaran1993experiments} & Active control of $(2,0)$ mode  & \\
			& \cite{thiessen2002active}  & in \cite{thiessen2002active} & \\\\
			Magnetic field  \hspace{1cm} &  \cite{nicolas1992magnetohydrodynamic} & Critical value of magnetic field\\\\
			Flow induced \hspace{1cm} &  \citep{lowry1997stability,lowry1994stabilization,lowry1995flow} & Axial flow\\\\
			Acoustic forcing &  \citep{marr1997stabilization,marr2001passive}& Radiation pressure \\\\
			Axial oscillation & \cite{chen1993nonlinear,mollot1993nonlinear}, & Axial oscillation of one disk\\ &\cite{benilov2016stability,haynes2018stabilization} & \\\\
			Radial forcing & \cite{patankar2018faraday} & Parametric stabilisation  & \\\\
			Electrochemical oxidation & \cite{song2020overcoming} & Controlling surface-tension  & 			
		\end{tabular}
		\caption{Literature on RP mode stabilisation} 
		\label{tab_lit}
	\end{center}
\end{table}

\section{Linear stability analysis}	
\begin{figure}
	\centering
	\includegraphics[scale=0.35]{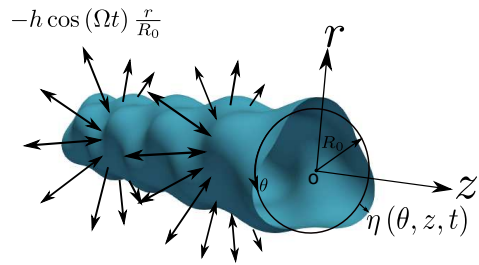}
	\caption{A cartoon of a surface perturbation on a viscous liquid cylinder of radius $R_0$ subject to a radial body force $\bm{\mathcal{F}}(r,t) = \mathcal{F}(r,t)\mathbf{\hat{e}}_r= -h\left(\frac{r}{R_0}\right)\cos(\Omega t)\mathbf{\hat{e}}_r$. The variable $\eta(\theta,z,t)$ measures the displacement of the free-surface with respect to the unperturbed cylinder, being zero in the base-state. Surface perturbations $\eta(\theta,z,t) = a_m(t;k)\cos(m\theta)\cos(kz)$ are imposed.}
	\label{fig1}
\end{figure}
Refer figure \ref{fig1}, the base-state comprises an infinitely long, quiescent liquid cylinder of density $\rho$, surface-tension $T$, kinematic viscosity $\nu$ and radius $R_0$ being subject to a radial, oscillatory body force $\mathcal{F}(r,t)$. This radial body force (per unit mass) has strength $h$ and a spatial dependence of the form $\frac{r}{R_0}$ in order to ensure single valuedness of the force at the origin \citep{adou2016faraday,patankar2018faraday} and the negative sign in the expression for $\mathcal{F}(r,t)$ is for convenience (see below equation \ref{eqn1}). Thus in the base state (variables with subscript $b$) there is no flow, the interface is a uniform cylinder of radius $R_0$ and the momentum equation simplifies to a balance between the radial oscillatory body force and the pressure gradient viz.
\begin{eqnarray}\label{eqn1}
		&\mathbf{u}_b = 0,\;\;-\frac{1}{\rho}\bm{\nabla}p_b + \mathcal{F}(r,t)\hat{\textbf{e}}_r = 0,\quad 0 \leq r \leq R_0  \\
		&\text{with} \quad \mathcal{F}(r,t) \equiv -h\left(\frac{r}{R_0}\right)\cos\left(\Omega t\right),\quad \text{and}\quad p_b(r,t) = \frac{\rho h}{2R_0}\left(R_0^2-r^2\right)\cos(\Omega t) + \frac{T}{R_0}. \nonumber
\end{eqnarray}
Here $\mathbf{\hat{e}}_r$ is the standard unit vector in the radial direction in cylindrical coordinates. Note that we have assumed stress in the fluid outside the cylinder to be zero, so that $p_b(R_0,t) = \frac{T}{R_0}$ satisfies the pressure jump condition at the interface due to surface tension. We neglect the density and viscosity of the fluid outside in the present study implying that the free-surface of the cylinder satisfies stress free conditions. In the following subsection, we briefly discuss RP modes in the unforced system ($h=0$) followed by inviscid and viscous description of RP stabilisation with radial forcing ($h\neq 0$).
\subsection{The inviscid and viscous RP modes ($h=0$)}
The classical RP modes are unstable axisymmetric Fourier modes satisfying $0 < kR_0 < 1$ for the unforced system ($h=0$). These are governed by the following inviscid (equation \ref{eqn2}a, \cite{rayleigh1878instability}) and viscous dispersion relation (\cite{rayleigh1892instability,weber1931zerfall,chandrasekhar1981hydrodynamic,liu2006linear}) with growth rate $\sigma_0$ (inviscid) and $\sigma$ (viscous) respectively.
\begin{subequations}\label{eqn2}
	\begin{gather}
	\sigma_0^2 = \frac{T}{\rho R_0^3}kR_0\left(1 - k^2R_0^2\right)\frac{\mi_1(kR_0)}{\mi_0(kR_0)},  \tag{\theequation a}\\	
	\sigma^{2}+2 v k^{2}\left[\frac{\mi_{1}^{\prime}(k R_{0})}{\mi_{0}(k R_{0})}-\frac{2 k l}{l^{2}+k^{2}} \frac{\mi_{1}(k R_{0})}{\mi_{0}(k R_{0})} \frac{\mi_{1}^{\prime}(l R_{0})}{\mi_{1}(l R_{0})}\right]\sigma
	- \left(\frac{l^{2}-k^{2}}{l^{2}+k^{2}}\right)\sigma_0^2 = 0,  \tag{\theequation b}\\
	\text{where}\quad l^2 \equiv k^2 + \frac{\sigma}{\nu} \nonumber
	\end{gather}	
\end{subequations}

\begin{figure}
	\centering
	\includegraphics[scale=0.22]{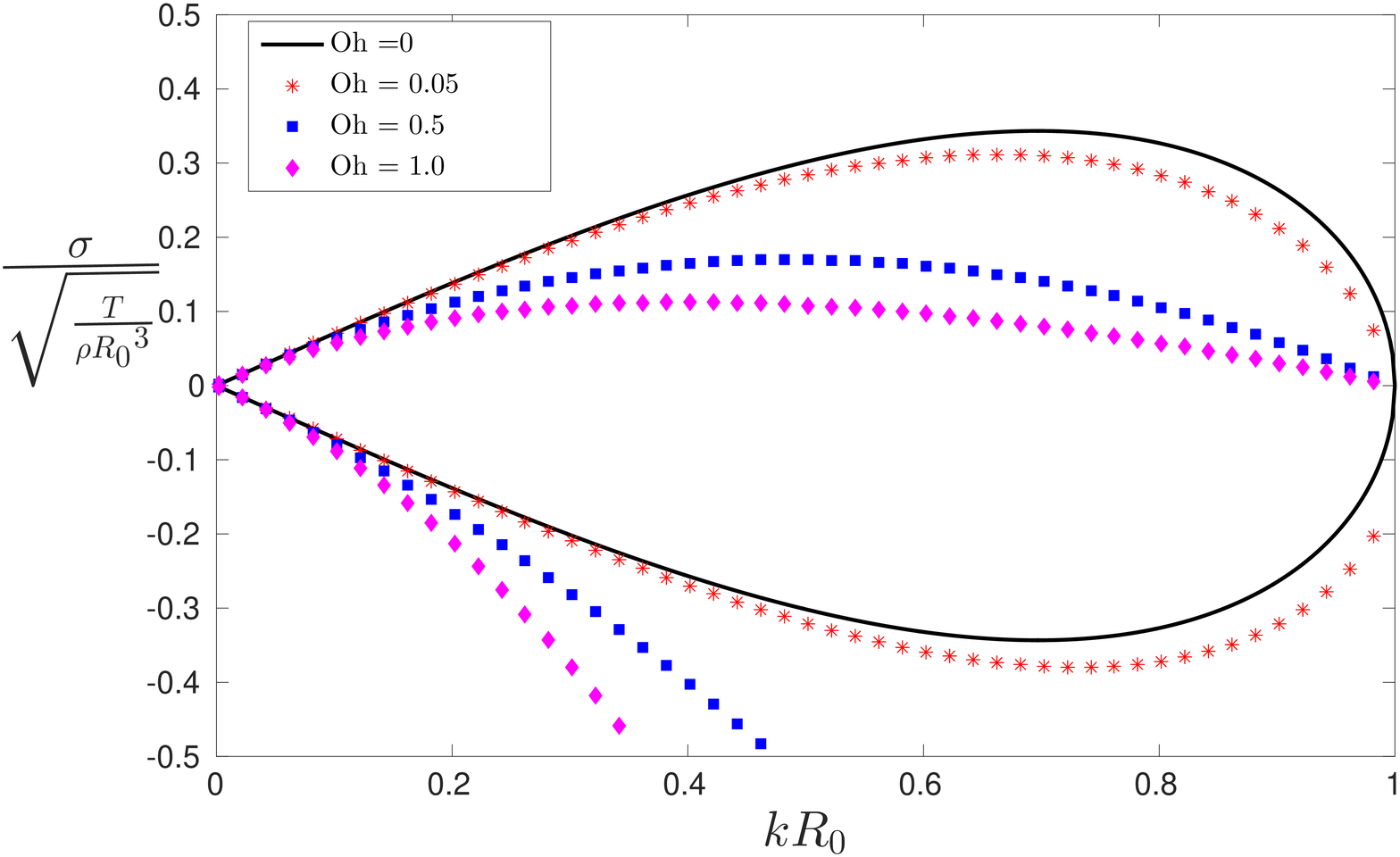}
	\caption{Inviscid and viscous growth (and decay) rates of RP modes ($0 < kR_0 < 1$) from numerically solving \ref{eqn2}a,b \citep{weber1931zerfall,garcia2008normal}. At any Ohnesorge (Oh) and $k$ in the range $0 < k < R_0^{-1}$, there are two capillary modes, one unstable ($\sigma > 0$) and another stable $(\sigma < 0)$. We stabilise the exponentially growing mode by forcing at $\Omega >> \sigma_{max}$ where $\sigma_{max}$ is the growth rate of the fastest growing RP mode, it being highest for the inviscid case ($Oh=0$) for $(kR_0)_{max} \approx 0.69$ with $\sigma_{\text{max}} \approx 0.34\sqrt{\frac{T}{\rho R_0^3}}$.}
	\label{fig2}	
\end{figure}
\noindent where $\mi_{m}(z)$ is the $m$th order modified Bessel function of the first kind and $\mi_{m}^{'}(z) \equiv \dfrac{d\mi_m}{dz}$. In figure \ref{fig2}, $\sigma_0$ and $\sigma$ are obtained by numerically solving eqns. \ref{eqn2}a and \ref{eqn2}b for the inviscid and viscous cases respectively. Unlike the inviscid relation \ref{eqn2}a which is quadratic in $\sigma_0$, the viscous dispersion relation given by \ref{eqn2}b  is transcendental in $\sigma$. It admits in addition to two capillary modes, a countably infinite set of hydrodynamic (or vorticity) modes as its roots and the latter are purely damped modes \citep{garcia2008normal}. In figure \ref{fig2} we only depict the growth and decay rates corresponding to the two capillary modes in the range $0 <kR_0 < 1$ for different values of Ohnesorge number $\text{Oh} = \frac{\mu}{\sqrt{T\rho R_0}}$. Our aim in this study is to stabilise the capillary modes in the range $0 < kR_0 < 1$ using radial forcing and this is discussed below.

\begin{figure}\label{fig3}. 
	\centering
	\subfloat[Stability plot]{\includegraphics[scale=0.24]{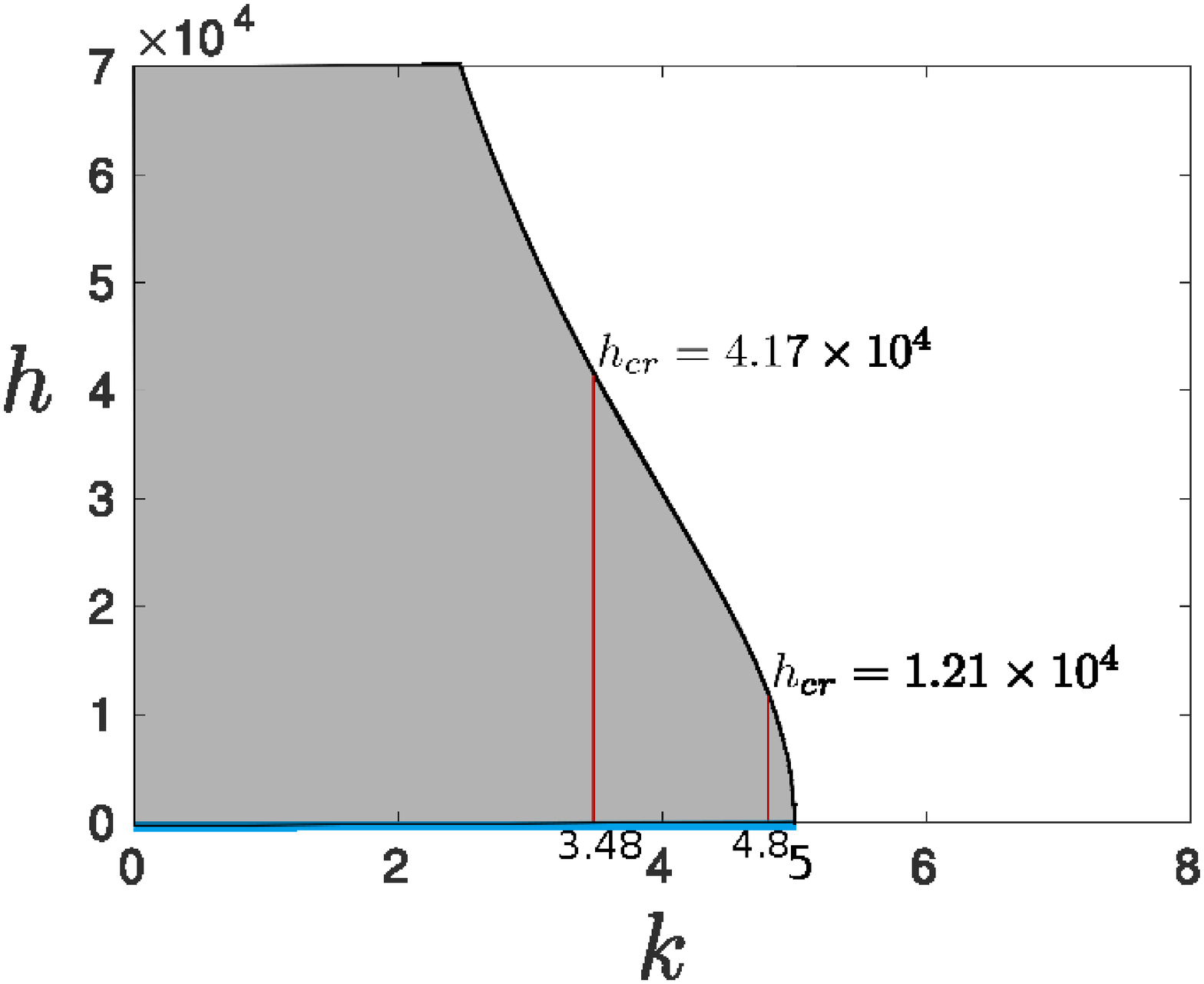}\label{fig3a}\hspace{2mm}}
	\subfloat[$k=4.8,h= 1.8\times 10^4$cm/s$^2$]{\includegraphics[scale=0.24]{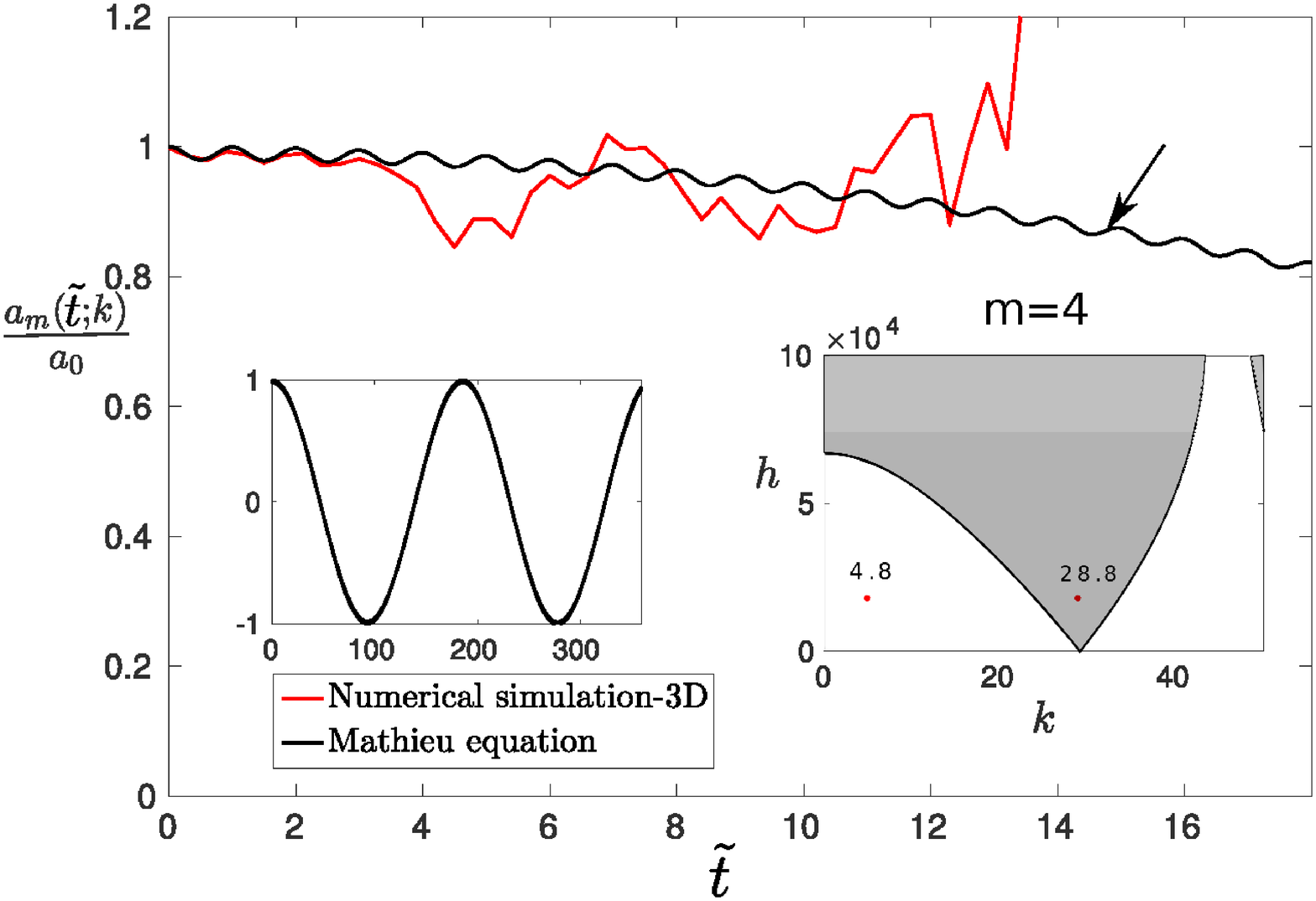}\label{fig3b}}
	\caption{Grey and white indicate unstable and stable regions respectively. \textbf{Panel (a)} Inviscid stablity chart for equation \ref{eqn3}. The forcing frequency $f = 300\;\text{Hz} >>$ $\sigma_{\text{max}}= 0.34\sqrt{T/(\rho R_0^3)} = 17.68\;$Hz. Parameters are for Case 1 in table \ref{tabparam2} with $\mu^{I}=0$. \\
	\textbf{Panel (b)} (Red curve) Time signal from numerical solution to the 3D Euler equation \citep{popinet2014basilisk} with an RP mode ($k_0=4.8\;$cm$^{-1},m_0=0$) excited at $t=0$. (Black curve) Solution to equation \ref{eqn3} (Left inset) Zoomed out view of solution to equation \ref{eqn3} (Right inset) Stability chart for $m=4$. An unstable non-axisymmetric Fourier mode ($k=28.8=6k_0,m=4$ in the grey region) at $\tilde{t} \approx 14$ s causes destabilisation of the cylinder.}
\end{figure}
\subsection{Dynamic stabilisation of RP modes - Linear inviscid theory}	
The inviscid results on RP stabilisation using radial forcing were presented earlier in \cite{patankar2018faraday} and are summarised very briefly here, for self-containedness. In the presence of radial forcing $\mathcal{F}(r,t) = -h\left(\frac{r}{R_0}\right)\cos(\Omega t)$  and under the linearised, inviscid, irrotational approximation, the equation governing the amplitude $a_m(t;k)$ of standing waves on the free surface of the form $\eta(z,\theta,t) = a_m(t;k)\cos(m\theta)\cos(kz)$ is the Mathieu equation \ref{eqn3} 
\begin{eqnarray}
\frac{d^2a_m}{dt^2} + \frac{\mi^{'}_{m}\left(kR_0\right)}{\mi_{m}\left(kR_0\right)}\left[\frac{T}{\rho R_0^3}kR_0\left(k^2R_0^2 + m^2 -1\right) + kh\cos\left(\Omega t\right)\right]a_m(t;k) = 0, \label{eqn3}
\end{eqnarray}
The stability diagram for equation \ref{eqn3} maybe obtained using Floquet analysis \citep{patankar2018faraday}. For $h \neq 0$, we have the interesting prediction that axisymmetric unstable RP modes can be stabilised by chosing $h$ to be sufficiently large. This is readily seen in the stability chart in figure \ref{fig3a} where the solid curve in black indicates the threshold value of forcing $h$ above which, a RP mode is stable. The line in blue indicates all unstable RP modes for $h=0$. Two representative RP unstable modes are chosen viz. $k_0=4.8$ cm$^{-1}$ (wavelength $\lambda \approx 1.309$ cm) and $k_0=3.48$ cm$^{-1}$ ($\lambda \approx 1.8$ cm) . The plot predicts the threshold values of forcing strength $h_{\text{cr}}=1.21\times 10^4$ cm/s$^2$ and $h_{\text{cr}}=4.17\times 10^4$ cm/s$^2$ respectively, beyond which these modes can be stabilised. For generating figure \ref{fig3a}, we have chosen $\Omega = 600 \pi$ rad/s (f=$300$ Hz), $R_0=0.2$ cm, density $\rho=0.957$ gm/cm$^3$, surface tension $T=20.7$ dynes/cm. These fluid parameters approximately correspond to silicone oil \citep{vega2009damping} with its viscosity artificially set to zero). Note that at these forcing frequencies, we may safely ignore compressibility effects as maybe inferred from the order of magnitude of the two typical velocity scales viz. $\text{maximum}\left[\frac{h_{c}}{f},fR_0\right] \approx 139$ cm/s for $f=300$Hz and $h_c=4.17\times10^4$cm/s$^2$. This is negligible compared to the typical acoustic speed $\mathcal{O}(10^5)$ cm/s in the fluid at ambient conditions. 

Figure \ref{fig3b} presents the time signal obtained from inviscid numerical simulations \citep{popinet2014basilisk} for the axisymmetric mode $k_0=4.8,m_0=0$ excited at $t=0$. Note that this is a RP unstable mode and as seen from figure \ref{fig3a}, it is expected to be stabilised beyond a threshold forcing of $h=1.21\times10^4$ cm/s$^2$. In fig. \ref{fig3b}, we see agreement between the solution to  equation \ref{eqn3} and the numerical simulation for very brief time (about three forcing time periods) after which the signal from the numerical simulation begins to deviate and grow rapidly (around $\tilde{t}\approx 14$) in contrast to the solution to equation \ref{eqn3} which stays bounded (see left inset). A Fourier analysis of the interface at $\tilde{t} \equiv t\Omega/2\pi \approx 14$ indicated by the arrow, reveals the appearance of a non-axisymmetric mode $(k=28.8,m=4)$ in the simulation. This is a stable mode in the unforced system ($h=0$) but is destabilised at the imposed level of forcing, lying inside a tongue as seen in the right inset of figure \ref{fig3b}. It becomes clear that for obtaining dynamic stabilisation, we need to ensure that all Fourier modes either present initially in the system or born via nonlinear effects, both axisymmetric and three-dimensional, should remain linearly stable at the imposed level of forcing. We will demonstrate in the next section that by taking viscosity into account and using the forcing frequency as a tuning parameter, this may be achieved.
	\subsection{Dynamic stabilisation of RP modes - Linear viscous theory}
Having demonstrated the inadequacy of dynamic stabilisation of RP modes in an inviscid model, we proceed to the viscous case. The motivation for including viscosity is simple to understand: it is known that inclusion of viscosity leads to displacement of the instability tongues upwards on the $h$-$k$ plane and these no longer touch the wavenumber axis \citep{kumar1994parametric}. Our expectation is that by suitably choosing viscosity and the forcing frequency, we will be able to shift the unstable tongues sufficiently above the wavenumber ($k$) axis. This generates a sufficiently large stable region where not only the axisymmetric RP unstable mode ($k_0$) is stablised (with forcing) but all higher modes accessible to the system are also stable. Note that the upward movement of the tongues occur not only for axisymmetric modes but also for non-axisymmetric ones. In particular we will also see that for fixed viscosity, we can move the minima of the tongue upwards by increasing the forcing frequency. The algebra for the viscous analysis is somewhat lengthy and details are provided in the supplementary material. We outline the important steps that follow. Expressing all quantities as sum of base plus perturbation i.e.
\begin{subequations}\label{eqn4}
	\begin{gather}
	\hat{p} = p_b + p,\quad \hat{\mathbf{u}} = \mathbf{0} + \mathbf{u}\;\;\; \&\;\;\text{perturbed free surface at}\; z = R_0 + \eta,  \tag{\theequation a,b,c}
	\end{gather}
\end{subequations}
Substituting \ref{eqn4}a,b into the incompressible Navier-Stokes equations and linearising about the base state we obtain the equations governing the perturbations viz.
\begin{subequations}\label{eqn5}
	\begin{gather}
	\left(\frac{\partial}{\partial t} - \nu \Delta\right)\mathbf{u} = -\frac{1}{\rho}\bm{\nabla}p,\;\;\mathbf{\nabla}\cdot\mathbf{u} = 0 \tag{\theequation a,b}
	\end{gather}
\end{subequations} 
where the vector Laplacian of the incompressible velocity field is $\Delta\mathbf{u} \equiv  -\bm{\nabla}\times\bm{\nabla}\times\mathbf{u}$.
The linearised boundary conditions are obtained by substituting \ref{eqn4}a,b,c into the boundary conditions (supplementary material), employing Taylor expansion and retaining terms linear in the perturbation variables viz. $\mathbf{u}, p$ and $\eta$ (the perturbation velocity $\mathbf{u}$ is written in terms of its components $(u_r,u_{\theta},u_z)$), we obtain 
\begin{subequations}\label{eqn6}
	\begin{align}
	&\frac{\partial\eta}{\partial t} = u_r(r=R_0), \tag{\theequation a}\\
	&\mu\left(\frac{\partial u_r}{\partial z} + \frac{\partial u_z}{\partial r}\right)_{r=R_0}  = 0, \quad
	\mu\left(r\frac{\partial}{\partial r}\left(\frac{u_{\theta}}{r}\right) + \frac{1}{r}\frac{\partial u_r}{\partial \theta}\right)_{r=R_0} = 0, \tag{\theequation b,c}\\	
	&\left( \frac{\partial}{\partial r} + \frac{1}{r}\right)\left[\frac{\partial u_r}{\partial t} - \nu\left\lbrace\Delta u_r - \frac{u_r}{r^2} - \frac{2}{r^2}\left(\frac{\partial u_{\theta}}{\partial\theta}\right)\right\rbrace\right] +\mathcal{F}(r,t)\Delta_O\eta - 2\nu\Delta_O\left(\frac{\partial u_r}{\partial r}\right) \nonumber \\
	&= -\frac{T}{\rho R_0^2}\Delta_O\left[\eta + \left(\frac{\partial^2\eta}{\partial \theta^2}\right) + R_0^2\left(\frac{\partial^2\eta}{\partial z^2}\right)\right] \;\;\text{at}\; r=R_0 \tag{\theequation d}, \\
	&\text{with}\quad\Delta_{O} \equiv \frac{1}{r^2}\frac{\partial^2}{\partial \theta^2} + \frac{\partial^2}{\partial z^2}, \nonumber \\
	& \mathbf{u}(r\rightarrow 0 ,t)\rightarrow \text{finite}. \tag{\theequation e} \\\nonumber
	\end{align}
\end{subequations} 
where $\Delta$ is the scalar Laplacian in cylindrical coordinates. Equations \ref{eqn6}a-e are the linearised versions of the kinematic boundary condition (equation \ref{eqn6}a), the zero shear stress condition(s) at the free surface (eqns. \ref{eqn6}(b,c)), the normal stress condition at the free-surface due to surface tension (equation \ref{eqn6}d) and the finiteness condition at the axis of the cylinder (equation \ref{eqn6}e) respectively. Eqn. \ref{eqn6}d has been obtained by eliminating pressure from the primitive form of pressure jump boundary condition (see supplementary material). Note the presence of the forcing term $\mathcal{F}(r,t)$ in the normal stress boundary condition in equation \ref{eqn6}d indicating the time periodicity of the base state.

We solve eqns. \ref{eqn5}a,b in the streamfunction-vorticity formulation and for this, the curl and double curl of equation \ref{eqn5}a leads to ($\bm{\omega} \equiv \bm{\nabla}\times \mathbf{u}$)
\begin{subequations}\label{eqn7}
	\begin{align}
	\frac{\partial\bm{\omega}}{\partial t} = \nu\bm{\Delta}\bm{\omega},\quad 
	\frac{\partial}{\partial t}\bm{\Delta} \mathbf{u} = \nu \bm{\Delta} \bm{\Delta} \mathbf{u}. \tag{\theequation a,b}
	\end{align}
\end{subequations}
where $\bm{\Delta}$ is the vector Laplacian. Employing the toroidal-poloidal decomposition \citep{marques1990boundary,boronski2007poloidal,prosperetti2011advanced}, the velocity and vorticity fields are expressed in terms of two scalar fields $\psi(r,\theta,z,t)$ and $\xi(r,\theta,z,t)$ using the decomposition
\begin{subequations}\label{eqn8}
	\begin{align}         	
	&\mathbf{u} = \bm{\nabla}\times\left(\psi\hat{\mathbf{e}}_z\right) + \bm{\nabla}\times\bm{\nabla}\times\left(\xi\hat{\mathbf{e}}_z\right),\quad \bm{\omega} \equiv \bm{\nabla}\times\bm{\nabla}\times\left(\psi\hat{\mathbf{e}}_z\right) + \bm{\nabla}\times\bm{\nabla}\times\bm{\nabla}\times\left(\xi\hat{\mathbf{e}}_z\right), \tag{\theequation a,b}
	\end{align}         	
\end{subequations} 
where $\hat{\mathbf{e}}_z$ is unit vector along the axial direction of the cylinder \citep{boronski2007poloidal}. By construction the velocity field in equation \ref{eqn8}a is divergence free and it can be shown (see supplementary material) that the equations governing the toroidal and poloidal fields $\psi(r,z,\theta,t)$ and $\xi(r,z,\theta,t)$ respectively, are the fourth and sixth order equations
\begin{subequations}\label{eqn9}
	\begin{gather}         	
	\left(\frac{\partial}{\partial t} - \nu\Delta\right)\Delta_{\text{H}}\psi = 0 \quad \text{and}\quad 
	\left(\frac{\partial}{\partial t} - \nu\Delta\right)\Delta\Delta_{\text{H}}\xi = 0, \tag{\theequation a,b}
	\end{gather}
\end{subequations}
where the scalar Laplacian $\Delta \equiv \frac{1}{r}\frac{\partial}{\partial r}\left(r\frac{\partial}{\partial r}\right) + \frac{1}{r^2}\frac{\partial^2}{\partial\theta^2} + \frac{\partial^2}{\partial z^2}= \Delta_{\text{H}} + \frac{\partial^2}{\partial z^2}$. 

As we have raised the order of our governing equations by taking curl and double curl, we need extra equations to determine the additional constants of integration. It was shown in \cite{marques1990boundary} that this takes the form of an additional equation also known as the compatibility condition \citep{boronski2007poloidal}. For the present problem at linear order, this extra equation is simply the radial component of the vorticity equation \ref{eqn7}a   \citep{boronski2007poloidal} i.e.
\begin{eqnarray} 
	&\dfrac{\partial \omega_r}{\partial t} = \nu\left\lbrace\Delta \omega_r - \dfrac{\omega_r}{r^2} - \dfrac{2}{r^2}\left(\dfrac{\partial \omega_{\theta}}{\partial\theta}\right)\right\rbrace \label{eqn10}\\
	&\text{with}\quad \omega_r = \dfrac{\partial^2 \psi}{\partial r \partial z }  -\dfrac{1}{r}\dfrac{\partial}{\partial\theta}\left(\Delta \xi \right) \quad \text{and} \quad \omega_\theta=\dfrac{1}{r}\dfrac{\partial^2  \psi}{\partial z \partial \theta } + \dfrac{\partial}{\partial r}\left(\Delta \xi\right) \nonumber
\end{eqnarray}          
In order to determine the scalar fields $\psi(r,\theta,z,t), \xi(r,\theta,z,t)$, we need to solve equations \ref{eqn9}(a,b). Analogous to the inviscid analysis in \cite{patankar2018faraday} we seek three dimensional standing wave solutions of the form 
\begin{subequations}\label{eqn11}
	\begin{align}
	&\psi(r,\theta,z,t) = \Psi_m(r,t;k)\sin(m\theta)\cos(kz),\quad\xi(r,\theta,z,t) = \Xi_m(r,t;k)\cos(m\theta)\sin(kz), \nonumber\\
	&\eta(\theta,z,t) = a_m(t;k)\cos(m\theta)\cos(kz),\tag{\theequation a,b,c} 
	\end{align}
\end{subequations}
where $k \in \mathbb{R}^{+}$ and $m\in \mathbb{Z}^{+}$. Substituting equations \ref{eqn11}(a,b) into eqns. \ref{eqn9} (a,b) we obtain the equations governing $\Psi_m(r,t;k)$ and $\Xi_m(r,t;k)$ viz.
\begin{subequations}\label{eqn12}
	\begin{align}
	&\left(\frac{\partial}{\partial t} - \nu \mathcal{L}\right)\mathcal{L_H}\Psi_m = 0,\quad \left(\frac{\partial}{\partial t} - \nu \mathcal{L}\right)\mathcal{L}\mathcal{L_H}\Xi_m = 0 \tag{\theequation a,b}\\
	&\text{where}\quad \mathcal{L_H} \equiv \frac{\partial^2}{\partial r^2} + \frac{1}{r}\frac{\partial}{\partial r} - \frac{m^2}{r^2} \;\;\& \quad \mathcal{L} \equiv \mathcal{L}_H  - k^2.\nonumber
	\end{align}
\end{subequations}    

Our task now is to determine the linear stability of the (time-dependent) base-state by identifying unstable and stable regions via Floquet analysis. This is indicated on the strength of forcing ($h$) versus wavenumber ($k$,$m$) plane for chosen fluid parameters $\rho,\nu, T$ and forcing frequency $\Omega$ and is done in the next subsection.
\subsubsection{Floquet analysis}
Using the Floquet ansatz for time periodic base states, we assume the following forms for $\Psi_m(r,t;k), \Xi_m(r,t;k)$ and $a_m(t;k)$ in equations \ref{eqn11}(a-c) \citep{kumar1994parametric}   
\begin{subequations}\label{eqn13}
	\begin{align}
	&\Psi_m(r,t;k) = \exp(\lambda_m(k) t)\sum_{n = -\infty}^{\infty}\tilde{\psi}_n^{(m)}(r;k)\exp(in\Omega t), \nonumber\\
	& \Xi_m(r,t;k) = \exp(\lambda_m(k) t)\sum_{n = -\infty}^{\infty}\tilde{\xi}_n^{(m)}(r;k)\exp(in\Omega t),\nonumber\\
	&a_m(t;k) = \exp(\lambda_m(k) t)\sum_{n = -\infty}^{\infty}\mathcal{M}_n\exp(in\Omega t), \nonumber 
	\tag{\theequation a,b,c}
	\end{align}
\end{subequations}
with $\lambda_m(k)$ being the Floquet exponent and $\tilde{\psi}_n^{(m)}(r;k)$ and $\tilde{\xi}_n^{(m)}(r;k)$ the complex eigenfunctions for each Fourier mode $(k,m)$. The complex eigenfunctions satisfy the reality condition $\tilde{\psi}_{-n}^{(m)} = \left(\tilde{\psi}_n^{(m)}\right)^{*}$ and $\tilde{\xi}_{-n}^{(m)} = \left(\tilde{\xi}_n^{(m)}\right)^{*}$, the superscript $^{*}$ indicating complex conjugation. 

We substitute \ref{eqn13}(a,b) into \ref{eqn12}(a,b) respectively yielding fourth and sixth order differential equations (eigenvalue problems) governing $\tilde{\psi}_n^{(m)}(r;k)$ and $\tilde{\xi}_n^{(m)}(r;k)$ for each $n$ in the expansion \ref{eqn13}(a,b)
\begin{subequations}\label{eqn14}
	\begin{align}
	&\mathbf{O}^{(k,m)}\cdot\left(\frac{d^2}{dr^2} + \frac{1}{r}\frac{d}{dr} - \frac{m^2}{r^2}\right)\tilde{\psi}_n^{(m)}(r;k) = 0, \tag{\theequation a}\\
	& \mathbf{O}^{(k,m)}\cdot\left(\frac{d^2}{dr^2} + \frac{1}{r}\frac{d}{dr} - \frac{m^2}{r^2}- k^2\right)\left(\frac{d^2}{dr^2} + \frac{1}{r}\frac{d}{dr} - \frac{m^2}{r^2}\right)\tilde{\xi}_n^{(m)}(r;k) = 0, \tag{\theequation b}
	\end{align}
\end{subequations}
where the linear operator $\mathbf{O}^{(k,m)} \equiv \left[\lambda_m(k) + in\Omega - \left(\frac{d^2}{dr^2} + \frac{1}{r}\frac{d}{dr} - \frac{m^2}{r^2}  - k^2\right)\right]$. Equations \ref{eqn14}(a,b) are solved with the finiteness condition at $r\rightarrow 0$ in equation \ref{eqn6}e leading to
\begin{subequations}\label{eqn15}
	\begin{align}      	
	\tilde{\psi}_n^{(m)}(r;k) = \mathcal{A}_n \mi_m(j_n r) + \mathcal{B}_n r^m,\quad\tilde{\xi}_n^{(m)}(r;k) = \mathcal{C}_n \mi_m(j_n r) + \mathcal{D}_n \mi_m(kr) +  \mathcal{E}_n r^{m} .\tag{\theequation a,b}
	\end{align} 
\end{subequations}
where $\mathcal{A}_n,\mathcal{B}_n,\mathcal{C}_n,\mathcal{D}_n$ and $\mathcal{E}_n$ are constants of integration,  $\mi_m(\cdot)$ is the $m^{\text{th}}$ order modified Bessel function of first kind and  $j_{n}^2 \equiv k^2 + \dfrac{\lambda_m(k) + in\Omega}{\nu}$ with $Re\{j_n\} > 0$. The compatibility condition in equation \ref{eqn10} may be further simplified using eqns. \ref{eqn11}(a,b), the Floquet ansatz \ref{eqn13}(a,b) and the expressions in \ref{eqn15}. The algebra for this is lengthy but eventually leads to a very simple relation viz.
\begin{eqnarray}\label{16} 
\mathcal{B}_n + k\mathcal{E}_n = 0\quad \forall\; n \;\in \mathbb{Z}.
\end{eqnarray} 
The constants $\mathcal{B}_n$ and $\mathcal{E}_n$ appear only in the combination $\mathcal{B}_n + k\mathcal{E}_n$ in subsequent algebra and thus equation \ref{16} may be used to eliminate these constants. Consequently the only constants which survive in further analysis are $\mathcal{A}_n,\mathcal{C}_n,\mathcal{D}_n$ and $\mathcal{M}_n$ (see equation \ref{eqn13}c). The Floquet ansatz in equation \ref{eqn13}(a,b) implies that the velocity components may be written as
\begin{eqnarray}\label{eqn17} 
\left(u_r,u_{\theta},u_z\right) = &\displaystyle\sum_{n=-\infty}^{\infty}\bigg(\tilde{u}_{r,n}(r)\cos(m\theta)\cos(kz),\tilde{u}_{\theta,n}(r)\sin(m\theta)\cos(kz),\tilde{u}_{z,n}(r)\cos(m\theta)\sin(kz)\bigg) \nonumber\\
&\times\exp\left[\left(in\Omega + \lambda_m(k)\right)t\right]
\end{eqnarray}
where the (complex) eigenmodes  $\tilde{u}_{r,n}(r),\tilde{u}_{\theta,n}(r)$ and $\tilde{u}_{z,n}(r)$ are determined using expressions \ref{eqn15}(a,b) in equations \ref{eqn8}a. These are
\begin{subequations}\label{eqn18}
	\begin{align}
	&\tilde{u}_{r,n}(r) =  \frac{m}{r}\mi_m(j_nr)\mathcal{A}_n + kj_n\mi_m^{'}(j_nr)\mathcal{C}_n + k^2\mi_m^{'}(kr)\mathcal{D}_n\nonumber \\ 
	&\tilde{u}_{\theta,n}(r) = -\bigg\{j_n \mi_m^{'}(j_n r)\mathcal{A}_n + \frac{km}{r}\bigg(\mi_m(j_nr)\mathcal{C}_n +  \mi_m(kr)\mathcal{D}_n\bigg)\bigg\} \nonumber \\
	& \tilde{u}_{z,n}(r) = -\bigg\lbrace j_n^2\mi_m(j_nr)\mathcal{C}_n + k^2\mi_m(kr)\mathcal{D}_n\bigg\rbrace, \tag{\theequation a,b,c}
	\end{align}
\end{subequations}
prime indicating differentiation with respect to the argument e.g. $\mi_m^{'}(z)\equiv \dfrac{d\mi_m}{dz}$ and so on. Note that despite the presence of terms of the form $1/r$ in expressions \ref{eqn18}(a,b), the velocity components do not diverge at the axis of the cylinder. This may be easily verified for the case $m > 0$ and the asymptotic form of $\mi_m(z)$ for small $z$.

The boundary conditions in eqns. \ref{eqn6}(a,b,c,d) may now be simplified employing expressions \ref{eqn17} and \ref{eqn18}(a,b,c) to obtain linear algebraic equations in $\mathcal{A}_n$, $\mathcal{C}_n,\mathcal{D}_n$ and $\mathcal{M}_n$. The algebra is provided in supplementary material and we provide only the normal stress boundary condition below
\begin{eqnarray}\label{eqn19}
& \Bigg[\mu\Bigg\{ k\mathcal{D}_n \left[ (k^2 - j_n^2)\dfrac{k\mi_m'(kR_0)}{R_0} - \left(k^2 + j_n^2 + \dfrac{2m^2}{R_0^2}\right)k^2\mi_m''(kR_0)\right] - 2  \left(k^2 + \frac{m^2}{R_0^2}\right)j_n^2\mi_m''(j_nR_0)k\mathcal{C}_n\nonumber\\
& -2\left(k^2 + \dfrac{m^2}{R_0^2}\right)\dfrac{m}{R_0}\left(j_n\mi_m'(j_nR_0) - \dfrac{\mi_m(j_nR_0)}{R_0}\right)\mathcal{A}_n \Bigg\} \nonumber \\
&- \dfrac{T}{R_0^2}\left(k^2 + \frac{m^2}{R_0^2}\right)\left(k^2R_0^2 + m^2-1\right)\mathcal{M}_n\Bigg]\left(\dfrac{2R_0^2}{\rho\left(k^2R_0^2 + m^2\right)}\right)   = h\left[\mathcal{M}_{n - 1} + \mathcal{M}_{n + 1}\right]  
\end{eqnarray}	

Equations \ref{eqn19} is solved symbolically in Mathematica using expressions for $\mathcal{A}_n$, $\mathcal{C}_n$ and $\mathcal{D}_n$ in terms of $\mathcal{M}_n$ to obtain a single equation relating $\mathcal{M}_{n-1},\;\mathcal{M}_{n}$ and $\mathcal{M}_{n+1}$ for $n=1,2,3\ldots N$. Equation \ref{eqn19} is thus written as a generalised eigenvalue problem
\begin{eqnarray}
\mathbf{A}\cdot\bm{\mathcal{M}} = h \; \mathbf{Q}\cdot\bm{\mathcal{M}}  \hspace{3cm} {n = 0,1,2,.....N}
\end{eqnarray}
where $\mathbf{A}$ and $\mathbf{Q}$ are matrices and we have taken $N=30$ terms in the Fourier series for this study (see supplementary material). Expressing  $\lambda_m(k) = \tilde{\mu}  + \mathup{I}\alpha$, the sub-harmonic case is $\alpha=\Omega/2$ and harmonic case is $(\alpha = 0)$ \citep{kumar1994parametric}. With $\tilde{\mu}=0$, the resultant equations are solved using the Matlab generalised eigenvalue solver \texttt{eig(,)},
MATLAB (2015) to obtain the stability boundaries on the wavenumber $k$ versus forcing $h$ plane for a given choice of $m$, forcing frequency $\Omega$ and fluid parameters $T,\rho,\mu$ and $R_0$. The stability charts obtained from Floquet analysis will be discussed in section \ref{sec_stab}.
\section{A non-local equation governing $\mathbf{a_m(t;k)}$}\label{sec_nle}
In this section, we present an analytical formulation which complements the Floquet analysis presented in section $2$. We obtain a self-contained equation for $a_m(t;k)$, the linearised amplitude of a Fourier mode $(\cos(kz),\cos(m\theta))$ in eqns. \ref{eqn11}(c). This equation will allow us to understand the physical role of viscosity. The starting point of the derivation are eqns. \ref{eqn12}(a,b). We define Laplace transforms as
\begin{eqnarray}\label{eqnIVP1}
\left[\tilde{\Psi}^{(m)}(r,s;k),\tilde{\Xi}^{(m)}(r,s;k),\tilde{a}_m(s;k)\right] = \int_{0}^{\infty}\exp\left(-st\right)\bigg[\Psi_{m}(r,t;k),\Xi_{m}(r,t;k),a_m(t;k)\bigg] dt \nonumber \\
\end{eqnarray}
In further algebra, the Laplace transform operator and its inverse are indicated as $\mathbf{\hat{L}}\left(\cdot\right)$ and $\mathbf{\hat{L}}^{-1}\left(\cdot\right)$ respectively and variables in the Laplace domain are indicated with a tilde on top. Laplace transforming equation \ref{eqn12}(a,b) with the initial conditions 
$\Psi_{m}(r,0;k) = \Xi_{m}(r,0;k) = 0$, $\dt{a}_m(0;k)=0$ and $a_m(0;k)=a(0)$ which correspond to deformation of the free surface and zero perturbation velocity (dot indicates time differentiation) initially, we obtain
\begin{subequations}\label{eqnIVP2}
	\begin{align}
	&\left(s - \nu \mathcal{L}\right)\mathcal{L}_H\tilde{\Psi}^{(m)}(r,s;k) = 0, \quad \left(s - \nu \mathcal{L}\right)\mathcal{L}\mathcal{L}_H\tilde{\Xi}^{(m)}(r,s;k) = 0 \tag{\theequation a,b}
	\end{align}
\end{subequations}         
The solution to equations \ref{eqnIVP2}(a,b) which stay finite as $r \rightarrow 0$ are the counterparts of expressions \ref{eqn15}(a,b). These are
\begin{subequations}\label{eqnIVP3}
	\begin{gather}          	
	\tilde{\Psi}^{(m)}(r,s;k) = \mathcal{A}(s)\mi_m(lr) + \mathcal{B}(s)r^m, \qquad \tilde{\Xi}^{(m)}(r,s) = \mathcal{C}(s)\mi_m(lr) + \mathcal{D}(s)\mi_m(kr) +  \mathcal{E}(s)r^m \\
	\textrm{where}\quad l^2(s) \equiv k^2 + \frac{s}{\nu}\;, \quad\mathit{Re}(l) > 0.\nonumber
	\end{gather}          
\end{subequations}
and $\mathcal{A}(s), \mathcal{B}(s), \mathcal{C}(s), \mathcal{D}(s)$ and $\mathcal{E}(s)$ are unknown functions to be determined subsequently. The algebra which follows is enormously simplified by recognising that the set of variables $\left[\mathcal{A}(s), \mathcal{B}(s), \mathcal{C}(s), \mathcal{D}(s),l^2\right]$ in this section are the analogues of the corresponding set $\left[\mathcal{A}_n, \mathcal{B}_n, \mathcal{C}_n, \mathcal{D}_n,j_n^2\right]$ used in the previous section.
The compatibility condition is thus
\begin{eqnarray}\label{eqnIVP4}
\mathcal{B}(s)+k\mathcal{E}(s)  = 0
\end{eqnarray}
and the normal stress boundary condition (equation \ref{eqn6}d) in the Laplace domain maybe written as 
\begin{eqnarray}\label{eqnIVP5}
\frac{T}{\rho R_0^2}\left(k^2R_0^2 + m^2-1\right)\tilde{a}_m &+& \frac{2\nu ml}{R_0}\mi_m^{'}(lR_0)\Lambda_2(s)\mathcal{A}(s) + 2\nu k l^2 \mi_m^{''}(lR_0)\mathcal{C}(s) \nonumber \\
&+& \left\lbrace2\nu k^3\mi_{m}^{''}(kR_0) + ks\mi_{m}(kR_0)\right\rbrace\mathcal{D}(s) - \mathcal{\tilde{F}}(R_0,s)\ast \tilde{a}_m(s;k) = 0 \nonumber \\
\end{eqnarray}		  
where the convolution term indicated with $*$ arises from the Laplace transform of the product of $\mathcal{F}(R_0,t)a_m(t;k)$ \citep{prosperetti2011advanced}. Analogous to the earlier section, from the other boundary conditions (equations \ref{eqn6}(a,b,c)) written in the Laplace domain we may obtain expressions for $\mathcal{A}(s), \mathcal{C}(s)$ and $\mathcal{D}(s)$ in terms of $\tilde{a}_m(s)$ and these are provided in Appendix A. These are substituted in  \ref{eqnIVP5} and produces the equation
\begin{eqnarray}
s\left(s\tilde{a}_m(s) - a(0)\right) &+& 2\nu k^2\frac{\mi_{m}^{''}(kR_0)}{\mi_{m}(kR_0)}\left(s\tilde{a}_m-a(0)\right) + 4\nu k\frac{\mi_{m}^{'}(kR_0)}{\mi_{m}(kR_0)}\tilde{\zeta}(s)\left(s\tilde{a}_m-a(0)\right) \nonumber \\
&+& \frac{\mi_m^{'}(kR_0)}{\mi_m(kR_0)}\tilde{\chi}(s)\bigg[\dfrac{T}{\rho R_0^3}kR_0\left(k^2R_0^2 + m^2-1\right)\tilde{a}_m - k\mathcal{\tilde{F}}(R_0,s)\ast\tilde{a}_m(s;k)\bigg] = 0 \nonumber \\\label{eqnIVP6}
\end{eqnarray}
where expressions for $\tilde{\chi}(s)$ and $\tilde{\zeta}(s)$ are provided below equation \ref{eqnIVP7}. Equation \ref{eqnIVP6} can be inverted into the time domain to obtain an integro-differential equation governing $a_m(t;k)$ (recall $\dt{a}_m(0;k)=0$)
\begin{eqnarray}
&\dfrac{d^2a_m}{dt^2} + 2\nu k^2\dfrac{\mathrm{I}_m''(kR_0)}{\mathrm{I}_m(kR_0)}\dfrac{da_m}{dt} + \displaystyle\int_{0}^{t}{\hat{\textbf{L}}}^{-1}\left(\tilde{\chi}(s)\right)\dfrac{\mathrm{I}_m'(kR_0)}{\mathrm{I}_m(kR_0)}\bigg[ \dfrac{T}{\rho R_0^3}kR_0\left(k^2R_0^2 + m^2-1\right)  \nonumber \\
&+ h k\cos\left[ \Omega (t - \tau) \right] \bigg] a_m(t - \tau) d\tau 
+ 4\nu k\dfrac{\mathrm{I}_m'(kR_0)}{\mathrm{I}_m(kR_0)}\displaystyle\int_{0}^{t}{\hat{\textbf{L}}}^{-1}\left[\zeta(s)\right]\dfrac{da_m}{dt}(t - \tau) d\tau = 0 \label{eqnIVP7}\\\nonumber \nonumber \\
&\text{where}\;
\tilde{\chi}(s) \equiv \dfrac{\left(k^2 - l^2\right)\Lambda_1(s) - 2k^2\Lambda_2(s) + 2l^2\Lambda_3}{2k^2\Lambda_2(s)-\left( l^2 + k^2 \right)\Lambda_1(s)},\nonumber\\
&\tilde{\zeta}(s) \equiv l\dfrac{\mathrm{I}_m'(lR_0)}{\mathrm{I}_m(lR_0)}\left\{\dfrac{2k^2\Lambda_2(s) - \left( l^2 + k^2 \right) \Lambda_3}{\left( l^2 + k^2 \right)\Lambda_1(s) - 2k^2\Lambda_2(s)}\right\}\Lambda_2(s) - k^2l\dfrac{\mathrm{I}_m''(lR_0)}{\mathrm{I}_m'(lR_0)}\left\{\dfrac{\Lambda_1(s) - \Lambda_3}{\left( l^2 + k^2 \right)\Lambda_1(s) - 2k^2\Lambda_2(s)}\right\},\nonumber 
\end{eqnarray}
while expressions for $\Lambda_1(s), \Lambda_2(s),\Lambda_3$ are provided in Appendix A. Note that since inversion of $\tilde{\chi}(s)$ and $\tilde{\zeta}(s)$ is not feasible analytically without further approximations, these inversions are indicated formally as $\hat{\textbf{L}}^{-1}(\cdot)$ in equation \ref{eqnIVP7}. Equation \ref{eqnIVP7} is one of the central results of our study and to the best of our knowledge this equation has not been derived in the literature before.

Equations \ref{eqnIVP6} and \ref{eqnIVP7} thus govern the amplitude of Fourier modes with indices $(k,m)$ in the Laplace and time domain respectively. These represent the cylindrical counterpart of the non-local equation governing viscous Faraday waves in Cartesian geometry, see  \citep{beyer1995faraday,cerda1997faraday}. The advantage of having an equation like \ref{eqnIVP7} for $a_m(t;k)$ is that it becomes possible to estimate separately, the viscous contributions to the time evolution of the free surface from damping in the irrotational part of the flow and from the boundary layer at the free-surface and this is done at the end of this study. We will demonstrate in section \ref{sec_dns} that the numerical solution to equation \ref{eqnIVP7} shows the stabilisation of RP modes that is sought and agrees very well with Direct Numerical Simulations. A number of consistency checks have been performed on equation \ref{eqnIVP6} and \ref{eqnIVP7} ensuring that these equations are consistent in various limits. These limits are discussed below.
\subsubsection*{Inviscid limit of equations \ref{eqnIVP6} and \ref{eqnIVP7}}
The first check on equation \ref{eqnIVP7} is to demonstrate that it reduces to equation \ref{eqn3} (Matheiu equation on an inviscid cylinder) in the inviscid limit. In the inviscid limit, $l\rightarrow \infty$ (for fixed $s$) and it maybe shown that $\lim_{\nu\rightarrow 0}\tilde{\zeta}(s)\rightarrow 0$ and $\lim_{\nu\rightarrow 0}\tilde{\chi}(s)\rightarrow 1$ in equation \ref{eqnIVP7}. For this, we have used the asymptotic expressions for $\mi_{m}(z)$ and $\mi_{m}^{'}(z)$ as $z\rightarrow \infty$ and fixed $m$ \citep{NIST:DLMF}.
Consequently the inversion of equation \ref{eqnIVP6} into the time domain becomes trivial leading to the Mathieu equation \citep{patankar2018faraday} for potential flow viz.
\begin{eqnarray}
&&\frac{d^2a_m}{dt^2} + \frac{\mathrm{I}_m'(kR_0)}{\mathrm{I}_m(kR_0)}\left[ \dfrac{T}{\rho R_0^3}kR_0\left(k^2R_0^2 + m^2-1\right) + kh \cos\left(\Omega t \right) \right] a_m(t) = 0
\label{eqnIVP8}
\end{eqnarray}
where we have used $\mathcal{F}(r,t) = -h\left(\frac{r}{R_0}\right)\cos(\Omega t)$ in writing equation \ref{eqnIVP8}.
%%%%%%%%%%%%%%%%%%%%%%%%%%%%%%%%%%%%%%%%%%%%%%%%%%
%%%%%%%%%%%%%%%%%%%%%%%%%%%%%%%%%%%%%%%   	   
\subsubsection*{Unforced ($h=0$) limit of equation \ref{eqnIVP6}}
The next test is to show that in the absence of forcing, expression \ref{eqnIVP6} leads to the correct dispersion relation for free, viscous modes. We demonstrate this for the axisymmetric case where expressions for $\tilde{\chi}(s)$ and $\tilde{\zeta}(s)$ (see below equation \ref{eqnIVP7}) are particularly very simple viz. for $m=0$, we have
\begin{eqnarray}
\tilde{\chi}(s) \rightarrow \frac{l^2-k^2}{l^2+k^2}=\frac{s}{s + 2\nu k^2},\;\; \tilde{\zeta}(s) \rightarrow -\frac{k^2l}{l^2+k^2}\frac{\mi_0^{''}(lR_0)}{\mi_0^{'}(lR_0)} = -\frac{\nu lk^2}{s + 2\nu k^2}\frac{\mi_0^{''}(lR_0)}{\mi_0^{'}(lR_0)} \label{eqnIVP9}     	  
\end{eqnarray}
These maybe obtained from the observation that for $m=0$, $\Lambda_1(s)$ diverges while $\Lambda_2(s)$ and $\Lambda_3$ remain finite. Using expressions \ref{eqnIVP9} in equation \ref{eqnIVP6} leads to,	
\begin{eqnarray}
	&&\left[s^2\Tilde{a}_0 - sa(0)\right] + 2\nu k^2\frac{\mathrm{I}_0''(kR_0)}{\mathrm{I}_0(kR_0)}\left[ s\Tilde{a}_0 - a(0)\right]  -4\nu k \frac{\mathrm{I}_0'(kR_0)}{\mathrm{I}_0(kR_0)}\frac{\nu lk^2}{s + 2\nu k^2}\frac{\mi_0^{''}(lR_0)}{\mi_0^{'}(lR_0)}\left[ s\Tilde{a}_0 - a(0)\right]\nonumber \\
	&&  + \frac{\mathrm{I}_0'(kR_0)}{\mathrm{I}_0(kR_0)}\frac{s}{(s + 2\nu k^2)}\left[ \frac{T}{\rho R_0^3} kR_0\left(k^2R_0^2-1\right) \Tilde{a}_0\right] = 0\label{eqnIVP10}     	  
\end{eqnarray}
implying
\begin{eqnarray}
\tilde{a}_0(s;k) = \dfrac{\left[s +2\nu k^2\dfrac{\mi_{0}^{''}(kR_0)}{\mi_{0}(kR_0)} - \dfrac{4\nu^2 l k^3}{s + 2\nu k^2}\dfrac{\mi_{0}^{'}(kR_0)}{\mi_{0}(kR_0)}\dfrac{\mi_0^{''}(lR_0)}{\mi_0^{'}(lR_0)} \right]}{s^2 + 2\nu k^2\Big\{\dfrac{\mi_{0}^{''}(kR_0)}{\mi_{0}(kR_0)} - \dfrac{2\nu l k}{s + 2\nu k^2}\dfrac{\mi_{0}^{'}(kR_0)}{\mi_{0}(kR_0)}\dfrac{\mi_0^{''}(lR_0)}{\mi_0^{'}(lR_0)}\Big\}s - \dfrac{s}{s + 2\nu k^2} \sigma_0^2 }a(0) \label{eqnIVP11}
\end{eqnarray}
Comparing the denominator of  equation \ref{eqnIVP11} with expression \ref{eqn2}b, and replacing $s\rightarrow\sigma$, we find that these are the same expressions. This is consistent as the viscous dispersion relation for free perturbations is obtained from the homogenous solution to the linear set of equations governing $\tilde{A}(s), \mathcal{\tilde{C}}(s),\mathcal{\tilde{D}}(s)$ and $\tilde{a}_m(s;k)$. The denominator of equation \ref{eqnIVP11} represents the determinant of the homogenous part of these equations \citep{prosperetti1976viscous,farsoiya_roy_dasgupta_2020} and thus leads us to the dispersion relation provided in equation \ref{eqn2}b. We have thus verified that equation \ref{eqnIVP6} produces the correct dispersion relation in the unforced, axisymmetric limit.
%%%%%%%%%%%%%%%%%%%%%%%%%%%%%%%%%%%%%%%
\subsubsection*{Flat interface limit of equation \ref{eqnIVP7}}
We demonstrate that in the limit $R_0\rightarrow\infty$ (flat interface limit), our equation \ref{eqnIVP6} reduces to the following equation ($\partial_t \equiv \frac{d}{dt}$) \citep{beyer1995faraday}
\begin{eqnarray}
	&&\bigg\{\frac{1}{k}{\left( \partial_t + 2\nu k^2 \right)}^2 + \left( \frac{Tk^2}{\rho} + h\cos\left(\Omega t \right) \right)\bigg\}a_0(t) \nonumber \\
	&&- \frac{4\nu^{3/2}k^2}{\pi}\int_{-\infty}^{t} \sqrt{\frac{\pi}{t - \tau}}\exp(-\nu k^2(t - \tau))\left(\partial_\tau + \nu k^2\right)a_0(\tau)d\tau = 0 \label{eqnIVP12} 
\end{eqnarray}
The algebra for this is lengthy and is provided in Appendix B. Equation \ref{eqnIVP12} is analogue of equation \ref{eqnIVP7} govering Faraday waves on a flat surface and was obtained by \cite{beyer1995faraday} (deep-water limit).  

Having demonstrated the consistency of equations \ref{eqnIVP6} and \ref{eqnIVP7}, we will return to analysing these at the end of section \ref{sec_dns}. Equation \ref{eqnIVP7} is solved numerically in Mathematica using built-in numerical Laplace inversion subroutines \citep{mathematica2017} and results will be compared with DNS in section \ref{sec_dns} in the context of RP stabilisation. In the next section, we discuss the stability plots obtained from Floquet analysis which will suggest the RP stabilisation strategy.  
\section{Linear stability predictions}\label{sec_stab}
\begin{figure} 
	\centering
	\subfloat[m=0]{%
		\includegraphics[scale=0.35]{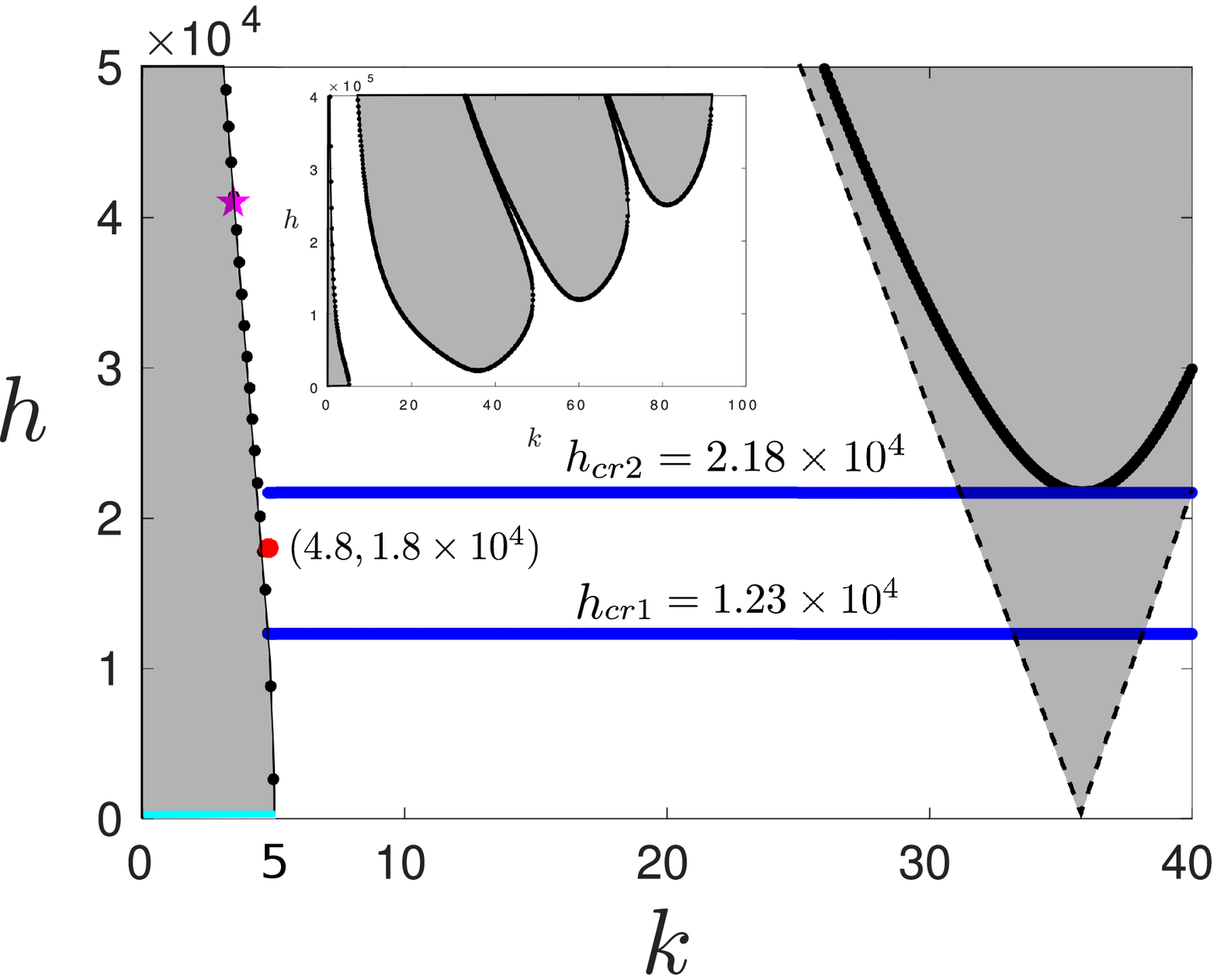}\label{fig4a}}\hspace{2mm}
	\subfloat[m=1,2,3,4]{%
		\includegraphics[scale=0.35]{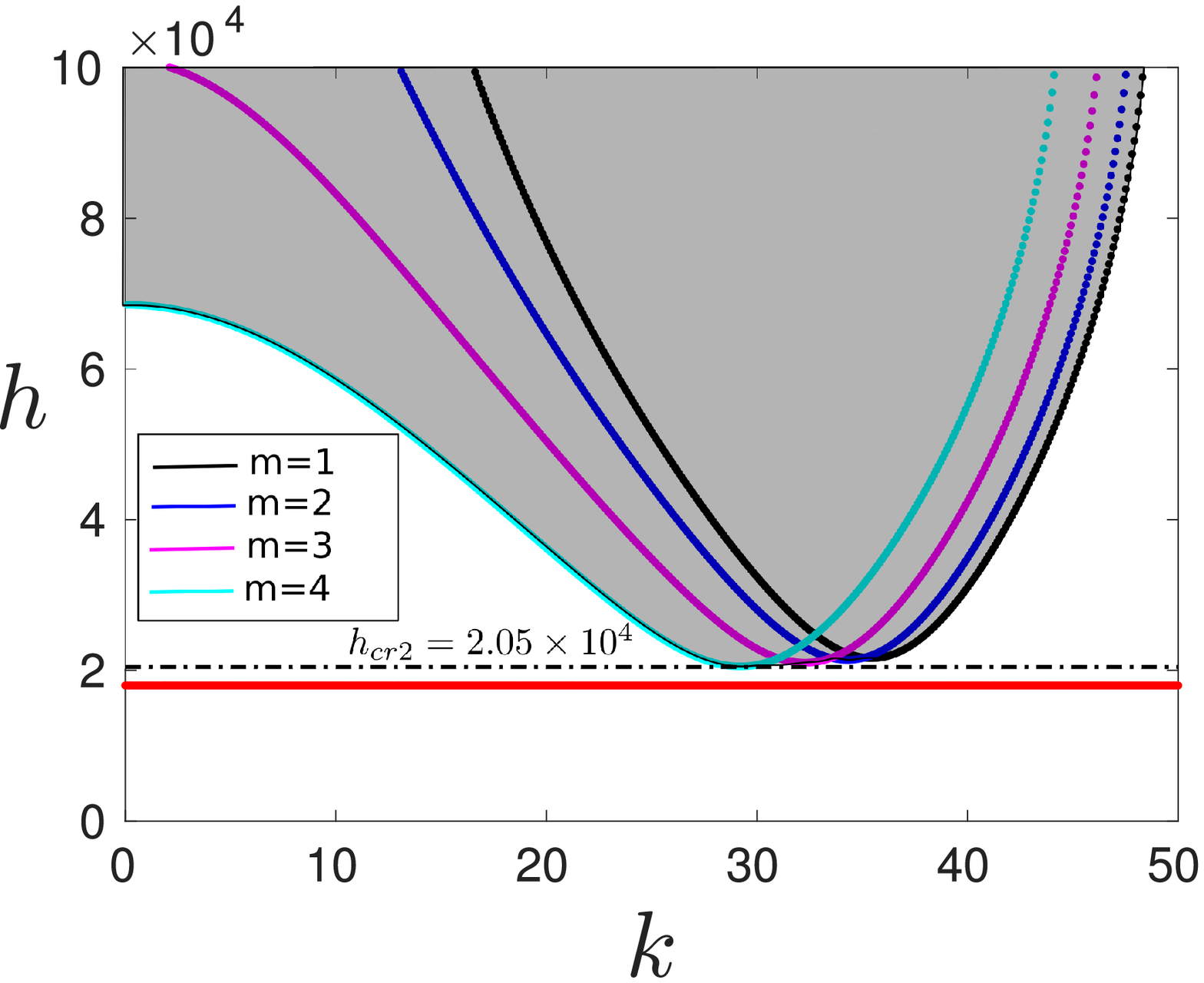}\label{fig4b}}
	\caption{\textbf{Panel a)} Stability plot for axisymmetric ($m=0$) and \textbf{panel (b)} non-axisymmetric ($m=1,2,3,4$) modes with Case 1 parameters, table \ref{tabparam2} ($\Omega = 600\pi$). For $h > 0$, grey and white regions are unstable and stable respectively. (Left panel) Bold black lines $\rightarrow$ viscous tongue, black dashed line$\rightarrow$ inviscid tongue. (Inset) de-magnified view. The mode ($k_0=4.8,m_0=0$) is stabilised for $h > h_{\text{cr1}}=1.23\times10^4$ cm/s$^2$. The optimum forcing satisfies $h_{\text{cr1}} < h <  h_{\text{cr2}}$ with $h_{\text{cr2}}=2.05\times 10^4$ cm/s$^2$ for $m=4$ (see right panel). The chosen $h=1.8\times10^4$ (indicated by red symbol and solid red line in left and right panels respectively) keeps the cylinder stable.}	
	\label{fig4}
\end{figure}

We discuss the stability plots on the $h$-$k$ plane obtained through Floquet analysis presented earlier.  
Refer to figure \ref{fig4a} (Case 1 in table \ref{tabparam2} provide the parameters), we wish to stabilise the axisymmetric RP unstable mode ($k_0=4.8,m_0=0)$ by subjecting the cylinder to an optimum forcing $h$. As shown in figure \ref{fig4a}, the viscous stability tongues are moved upwards due to viscosity \citep{kumar1994parametric}, compared to the inviscid tongues which touch the wavenumber axis (black dashed line in left panel). The figure shows that the critical threshold of forcing (we will call it $h_{\text{cr1}}$ hereafter) for stabilising ($k_0=4.8,m_0=0)$ is $h_{\text{cr1}}=1.23\times 10^4$ cm/s$^2$, and the applied forcing ($h$) needs to satisfy $h > h_{\text{cr1}}$ for stabilisation of this mode. Simultaneously, we also need to ensure that $h$ is below a second threshold $h_{\text{cr2}}$. This second threshold ($h_{\text{cr2}}$) is chosen to be the ordinate corresponding to the lowest minima among all the stability tongues in figs. \ref{fig4a} and \ref{fig4b}. For stabilisation we require $h_{\text{cr1}} < h_{\text{cr2}}$ and this is ensured by using the frequency of forcing $\Omega$ as a control parameter for a given set of fluid parameters. Once we have chosen an $\Omega$ which satisfies the ordering $h_{\text{cr1}} < h_{\text{cr2}}$, any choice of $h$ satisfying $h_{\text{cr1}} < h < h_{\text{cr2}}$ not only stabilises the primary mode ($k_0,m_0$) but also keeps moderately high modes ($k > k_0$ for $m=0,1,2,3,4\ldots$) stable. 

Note that viscosity plays a very important role in this stabilisation as by displacing the (in)stability tongues upward, it allows for the possibility of choosing the forcing such that $h_{\text{cr1}} < h < h_{\text{cr2}}$. In the inviscid case, this is impossible to arrange as $h_{\text{cr}2}=0$ because in the inviscid case all (instability) tongues touch the wavenumber axis. Consequently in an inviscid system if we force the cylinder at $h > h_{\text{cr}1}$, while the RP mode ($k_0,m_0=0$) is definitely stabilised, at long time \citep{patankar2018faraday} higher modes (axisymmetric and non-axismmetric) are produced due to nonlinearity and some of these are inevitably linearly unstable at the chosen level of forcing $h$. As a consequence, the stabilisation in inviscid systems in short-lived thus rendering the stabilisation strategy unsuitable (this was shown in figure \ref{fig3b}). The situation is rectified by including viscosity into our analysis. Refer to figure \ref{fig4} where the red dot in the left panel and the solid red line in the right panel indicates a suggested optimal value of $h$ satisfying $h_{\text{cr1}} < h < h_{\text{cr2}}$ for the RP mode $k_0=4.8,m_0=0$. Note that the high modes (i.e. those with $k >> k_0$ and $m>>m_0$) which can be generated due to nolinearity, are also associated with high rates of dissipation. Consequently we need not take into account the stability of very high modes in our stabilisation strategy. For the present purpose, we found it adequate to ensure that at the chosen value of $\Omega$ and $h$, the primary mode $(k_0,m_0)$ as well as modes upto $(7k_0,\;m=0,1,2,3,4)$ are stable. This is found to be adequate for stabilisation of the liquid cylinder for several forcing time-periods. 

An important point to note here is that although our theory has been developed assuming that a continuous range of RP modes with arbitrary long wavelengths ($k \rightarrow 0$) are accessible to our system, in practise there is a finite upper limit on the maximum wavelength that the system can access (due to axial confinement). In validating the present stability predictions via direct numerical simulations (see section \ref{sec_dns}), we chose the length $L$ of the unperturbed cylinder to be $L=\frac{2\pi}{k_0}$, $k_0$ being the wavenumber of the axisymmetric RP unstable mode we intend to stabilise. Boundary conditions (periodic) in the axial ($z$) direction imply that only integral multiples of wavenumber $k_0$ are allowed to appear in our simulations. This ensures that wavenumbers verifying $k < k_0$ are not accessible to our system, although it is clear from figure \ref{fig4a} that such axisymmetric modes can continue to be unstable at the optimal level of forcing ($h=1.8\times10^4$). We shall return to this point at the end of this study. For stabilising the mode ($k_0=4.8,m_0=0$), we have chosen $h=1.8\times10^4$ (satisfying $h_{\text{cr}1} < h < h_{\text{cr}2}$) as indicated by the red dot in figure \ref{fig4a}. It will be shown in section \ref{sec_dns} through direct numerical simulations (DNS) that exciting the perturbation $k_0=4.8,m_0=0$ on the cylinder at $t=0$ with the forcing strength $h=1.23\times10^4$ (at $\Omega=600\pi$), allows it to remain stable upto several hundred forcing time periods. The imposed perturbation decays to zero at long time, in excellent agreement with the solution to equation \ref{eqnIVP7}. 
\begin{figure}
	\centering
	\subfloat[m=0]{%
		\includegraphics[scale=0.26]{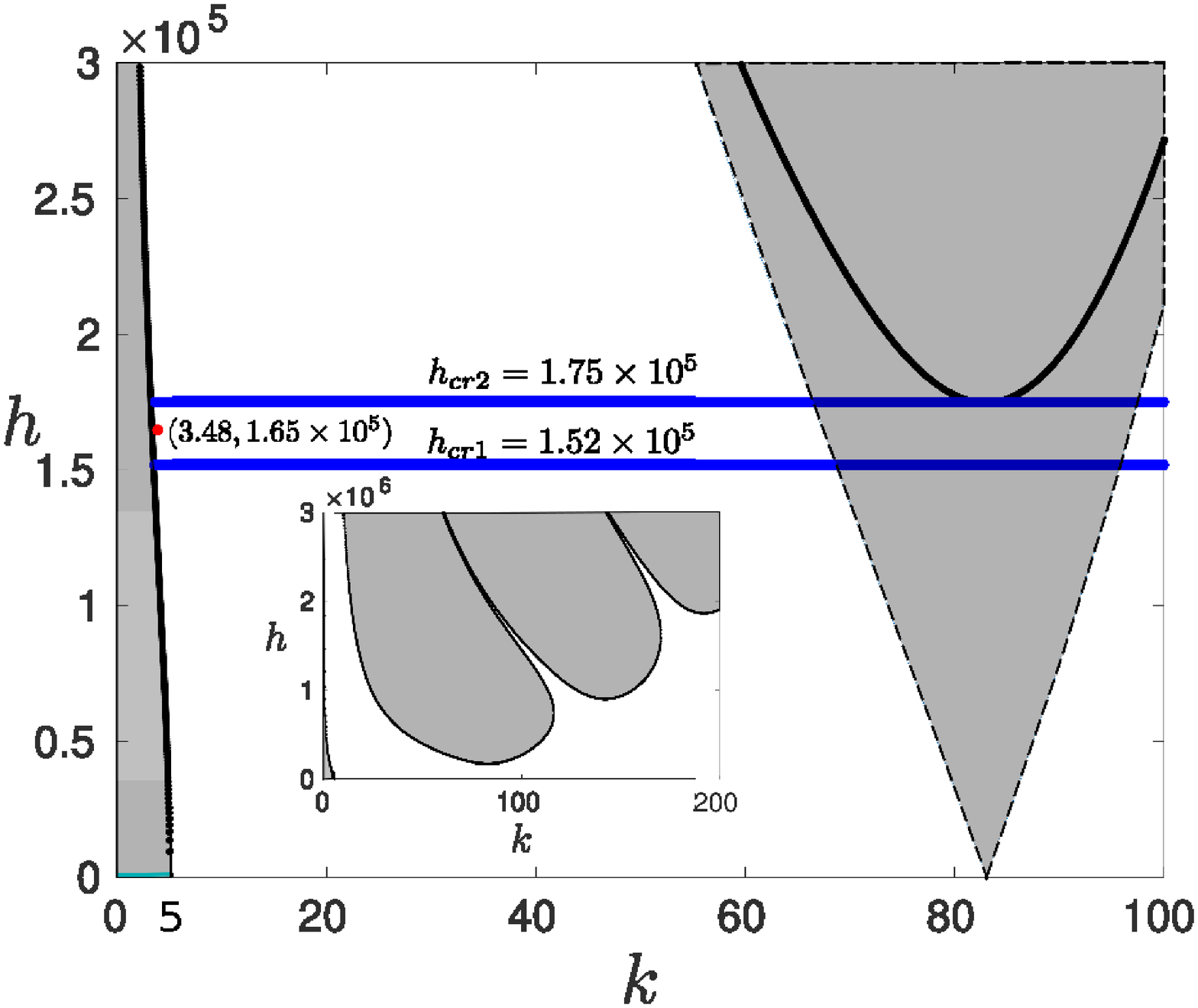}\label{fig5a}}\hspace{4mm}
	\subfloat[m=1,2,3,4]{%
		\includegraphics[scale=0.26]{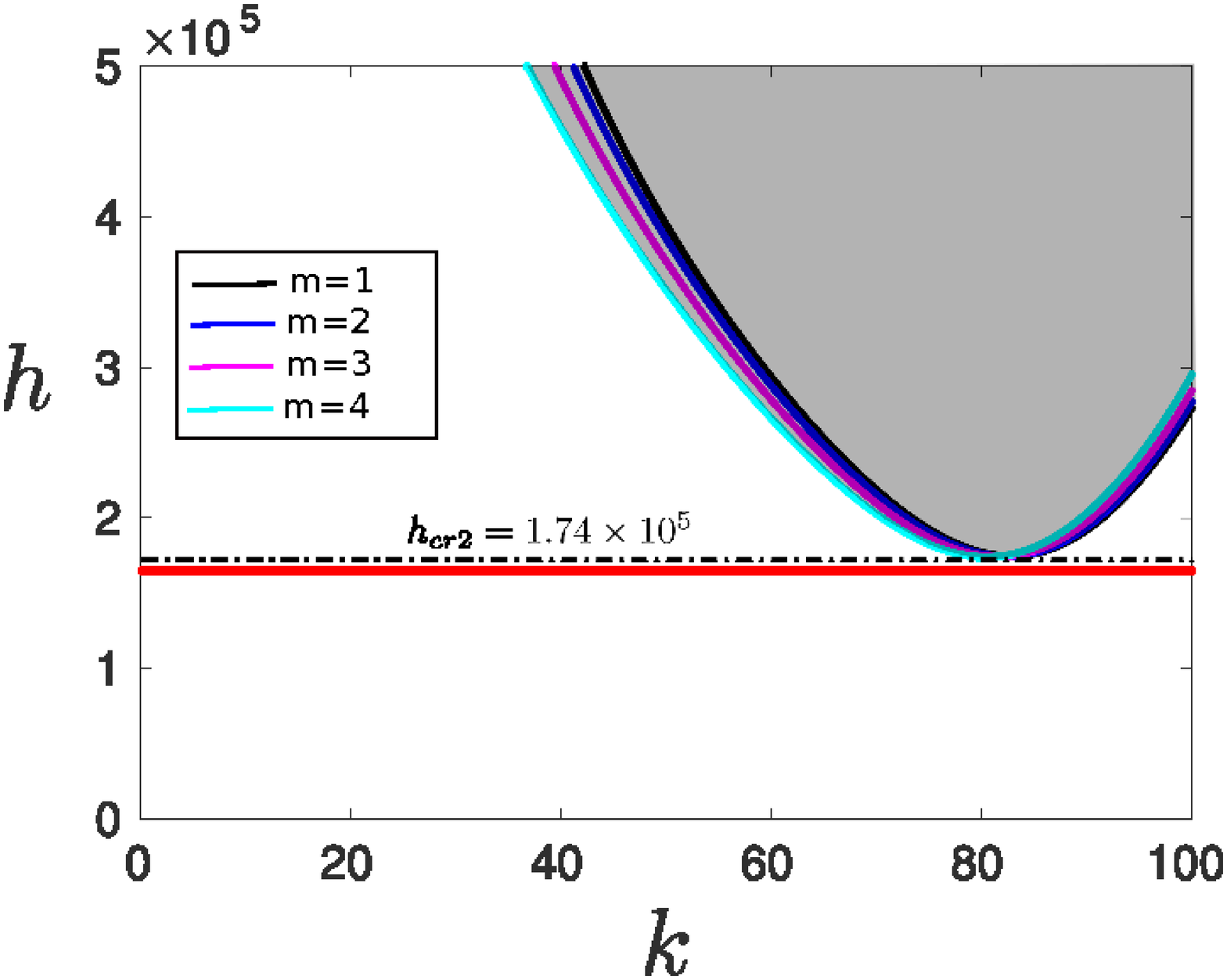}\label{fig5b}}
	\caption{\textbf{Panel a)} Stability plot for axisymmetric $m=0$ and \textbf{panel b)} non-axisymmetric ($m=1,2,3,4$) modes with case 2 parameters, table \ref{tabparam2} ($\Omega=2200\pi$). The figures are to be read in the same way as figure \ref{fig4}. The mode ($k_0=3.48,m_0=0$) is stabilised for $h > h_{\text{cr1}}=1.52\times10^5$ cm/s$^2$. The optimum forcing satisfies $h_{\text{cr1}} < h <  h_{\text{cr2}}$. Here $h_{\text{cr2}}=1.74\times 10^5$ cm/s$^2$ for $m=4$ (right panel). The chosen $h=1.65\times10^5$ (indicated by red symbol and solid red line in left and right panel respectively) keeps the cylinder stable.}
	\label{fig5}			
\end{figure}

We next provide the optimal forcing strength for a slightly longer wavelength RP mode compared to the previous case. We choose to stabilise the axisymmetric RP unstable mode $(k_0=3.48,m_0=0)$. This mode is indicated with a pink star in figure \ref{fig4a}. It is seen that $h_{\text{cr}1}$ for this mode is $\approx 4.1\times 10^4$ cm/s$^2$ and thus we do \textit{not} satisfy $h_{\text{cr1}} < h_{\text{cr2}}$ (the minima of all the axisymmetric and non-axisymmetric stability tongues are much lower than $h_{\text{cr1}}$). Choosing simply $h > h_{\text{cr}1}$ allows the possibility of higher unstable modes to appear in simulations, as discussed in the last paragraph. In order to prevent this we now use the forcing frequency $\Omega$ as a tuning parameter. In figure \ref{fig5}, we have increased $\Omega=2200\pi$ (from $600\pi$ earlier) holding all fluid parameters at the same value as earlier (this is Case $2$ in table \ref{tabparam2}). The advantage of doing so is visible in figs. \ref{fig5a} and  \ref{fig5b} where it is seen that by increasing $\Omega$, we have the desired ordering. For our chosen mode $(k_0=3.48,m_0=0)$, we can see that $h_{\text{cr}1} \approx 1.52\times 10^5$ and $h_{\text{cr}2} \approx 1.74\times 10^5$ (obtained from the minima of the $m=4$ tongue shown in the right pane) and the desired ordering $h_{\text{cr1}} < h_{\text{cr2}}$ exists at this forcing frequency. The optimal level of forcing is chosen to be $h=1.65\times10^5$ cm/s$^2$ (indicated by the red dot and the solid red line in the left and right pane respectively). It will be shown in the next section through DNS that this mode is also stabilised at this optimal forcing for more than two thousand forcing time periods.

\section{Numerical simulations}\label{sec_dns}
We compare the predictions made in the previous section(s) with direct numerical simulations (DNS). The simulations are executed using Basilisk \citep{popinet2014basilisk} which solves the incompressible, Navier-Stokes equations for two-fluids with outer fluid density and viscosity $\rho^{\mathcal{O}}, \mu^{\mathcal{O}}$ and inner fluid parameters $\rho^{\mathcal{I}}, \mu^{\mathcal{I}}$. As our theory neglects the outer fluid, the ratios $\rho^{\mathcal{O}}/\rho^{\mathcal{I}}$ and $\mu^{\mathcal{O}}/\mu^{\mathcal{I}}$ have both been chosen to be quite small to minimise the dynamics of the outer fluid. Basilisk is based on the Volume of Fluid (VoF) algorithm and the solver has been extensively benchmarked for unsteady two-phase flows \citep{farsoiya2021bubble,basak_farsoiya_dasgupta_2021,mostert_deike_2020,singh2019test,farsoiya2017axisymmetric}. A comprehensive list of publications based on the Basilisk solver is provided in \cite{popinet2014basilisk}.

\begin{figure}
	\centering
	\includegraphics[scale=0.2]{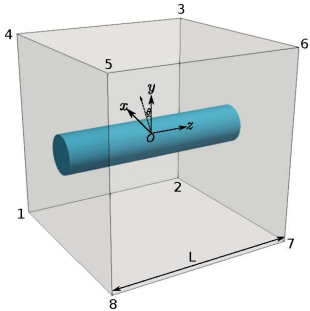}\label{}
	\caption{DNS geometry. A radial body force $\bm{\mathcal{F}}(r,t) =  -h\left(\frac{r}{R_0}\right)\cos\left(\Omega t\right)\mathbf{\hat{e}}_r$ is applied at every grid point in the domain. Boundary conditions are listed in table \ref{tab_bc}. The length of the domain $L=\frac{2\pi}{k_0}$, $k_0$ being the wavenumeber of the axisymmetric RP unstable mode that is excited at $t=0$}
	\label{fig6}			
\end{figure}

The computational geometry and the boundary condtions are shown in figure \ref{fig6} and table \ref{tab_bc} respectively. For numerical reasons we have applied the radial forcing term $\bm{\mathcal{F}}(r,t) =  -h\left(\frac{r}{R_0}\right)\cos\left(\Omega t\right)\mathbf{\hat{e}}_r$ to the entire computational domain in figure \ref{fig6}. As the density of the outer fluid is very small (viz. $\rho^{\mathcal{I}}/\rho^{\mathcal{O}} \approx 10^3$), the effect of forcing on the outer fluid remains small and results from the DNS will be seen to agree very well with theory which ignores the effect of the outer fluid.
%%%%%%%%%%%%%%%%%%%%%%%%%%%%
\begin{table}
	\begin{center}
		\begin{tabular}{lcccc}
			\textbf{Sl.} & \textbf{Face} & \textbf{Pressure} $(p)$ & \textbf{Velocity} $(u, v, w)$ & \textbf{Volume fraction} $(c)$ \\
			1 & $1854$, $2763$ & Periodic & Periodic & Periodic \\
			2 & $1234,5678,3456,1278$ & Dirichlet & Neumann & Neumann \\
		\end{tabular}
		\caption{ Boundary conditions for 3D DNS.}
		\label{tab_bc}
	\end{center}
\end{table}
A base level refinement of $6$ (in powers of two) with adaptive higher grid levels of $9$ are employed at the interface and for fluid inside the cylinder. Table \ref{tab_bc} lists the boundary conditions used on the various faces of the domain. Note that for axisymmetric simulations, we use symmetry conditions on the axis of the cylinder. The length of the computational domain is $L=\frac{2\pi}{k_0}$ where $k_0$ is the RP unstable mode we wish to stabilise. The interface is deformed initially as  $\eta(z,\theta,0) = a_m(0)\cos(k_0z)$ with zero velocity everywhere in the domain and we track the evolution of the interface with time at the centre of the domain (see figure \ref{fig6}). Baslisk \citep{popinet2014basilisk} solves the following equations
\begin{eqnarray}
&&\frac{D\bm{u}}{Dt} = \rho^{-1} \left\{ - \bm{\nabla }p + \bm{\nabla }\cdot (2\mu \mathbf{D}) + T \kappa \delta_s \mathbf{n} \right\} - h\cos(\Omega t)\frac{r}{R_0}\mathbf{e}_r, \label{ch2ns}\\
&& \bm{\nabla} \cdot \bm{u} = 0 \quad\text{and}\quad \frac{\partial c}{\partial t} + \bm{\nabla}\cdot(c\bm{u})=0, \label{ch2cont_vof}
\end{eqnarray}    
where $\rho \equiv c\rho^{\mathcal{I}} + (1-c)\rho^{\mathcal{O}}$, $\mu \equiv c\mu^{\mathcal{I}} + (1-c)\mu^{\mathcal{O}}$, $\bm{u}$, $p$, $\mathbf{D} = [\bm{\nabla }\bm{u } + (\bm{\nabla }\bm{u })^{Tr}]/2$, $c$ are density, velocity, pressure, stress tensor and volume fraction respectively. The volume fraction field $c$ is unity for fluid inside the filament and $0$ for the fluid outside. $T$ is the surface tension coefficient, $\delta_s$ is a surface delta function, $\kappa \equiv \frac{1}{\mathcal{R}}$ is the local curvature, $\mathbf{n}$ is a local unit normal to the interface and $R_0$ is the radius of the unperturbed filament.

%%%%%%%%%%%%%%%%%%%%%%%%%%%%
\begin{table}
	\centering
	\fontsize{6.8}{6}\selectfont

	\begin{tabular}{P{0.7cm}P{1.5cm}P{0.4cm}P{0.4cm}P{0.5cm}P{0.5cm}P{0.7cm}P{0.5cm}P{0.7cm}P{0.5cm}P{1.3cm}P{0.7cm}P{0.5cm}P{0.1cm}}
		\toprule		
		\textbf{Case} & \textbf{Fluid} & $a(0)$ & $m_0$ &  $k_0$ & $\rho^{\mathcal{I}}$ & $\rho^{\mathcal{O}}$ &$\mu^{\mathcal{I}}$&$\mu^{\mathcal{O}}$& $R_0$ & $h$ & $\Omega$ & $T$ & \\ 
		\midrule[0.35mm]		
		\midrule
		1 & properties close to silicone oil & $0.01$ & $0$&  $4.8$ & $0.957$ & $0.001$ & $0.1$ & $0.001$ & $0.2$ & $1.8\times10^4$ &$600\pi$ & $20.7$ & \\	
		\midrule
		2 & -do- & $0.01$ & $0$&  $3.48$ & $0.957$ & $0.001$ & $0.1$ & $0.001$ & $0.2$ & $1.65\times10^5$ &$2200\pi$ & $20.7$ & \\	
		\midrule
		3 & -do- & $0.01$ & $0$&  $4.8$ & $0.957$ & $0.001$ & $0.2$ & $0.001$ & $0.2$ & $1.8\times10^4$ &$600\pi$ & $20.7$ & \\		
		\midrule
	\end{tabular}
	\caption{\textbf{DNS Parameters} (CGS units)}
	\label{tabparam2}
\end{table}
\begin{figure}
	\centering
	\includegraphics[scale=0.28]{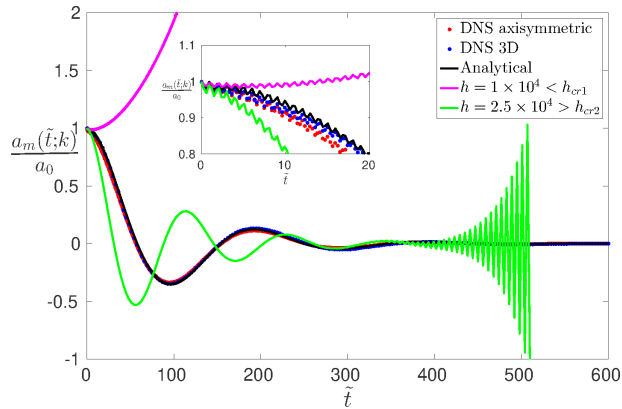}
	\caption{Case $1$ in Table \ref{tabparam2}: (Red and blue dots) DNS time signal for ($k_0=4.8,m_0=0$) excited at $t=0$ and $h_{\text{cr1}} < h < h_{\text{cr}2}$, refer stability plot in figure \ref{fig4}. (Black line) Solution to equation \ref{eqnIVP7} (Pink line) Destabilisation seen in axisymmetric DNS when $h < h_{\text{cr}1}$ and when (Green line) $h > h_{\text{cr}2}$. Note the excellent agreement between solution to equation \ref{eqnIVP7} and DNS upto $600$ forcing cycles ($\tilde{t} \equiv t\Omega/2\pi$). This is in contrast to inviscid simulations in figure \ref{fig3b} where for the same $k_0$, stabilisation is seen for only three forcing cycles.}
	\label{fig7}
\end{figure}
%%%%%%%%%%%%%%%%%%%%%%%%%%%%
\begin{figure}
	\centering
	\subfloat[Case 1 in table \ref{tabparam2}]{%
		\includegraphics[scale=0.22]{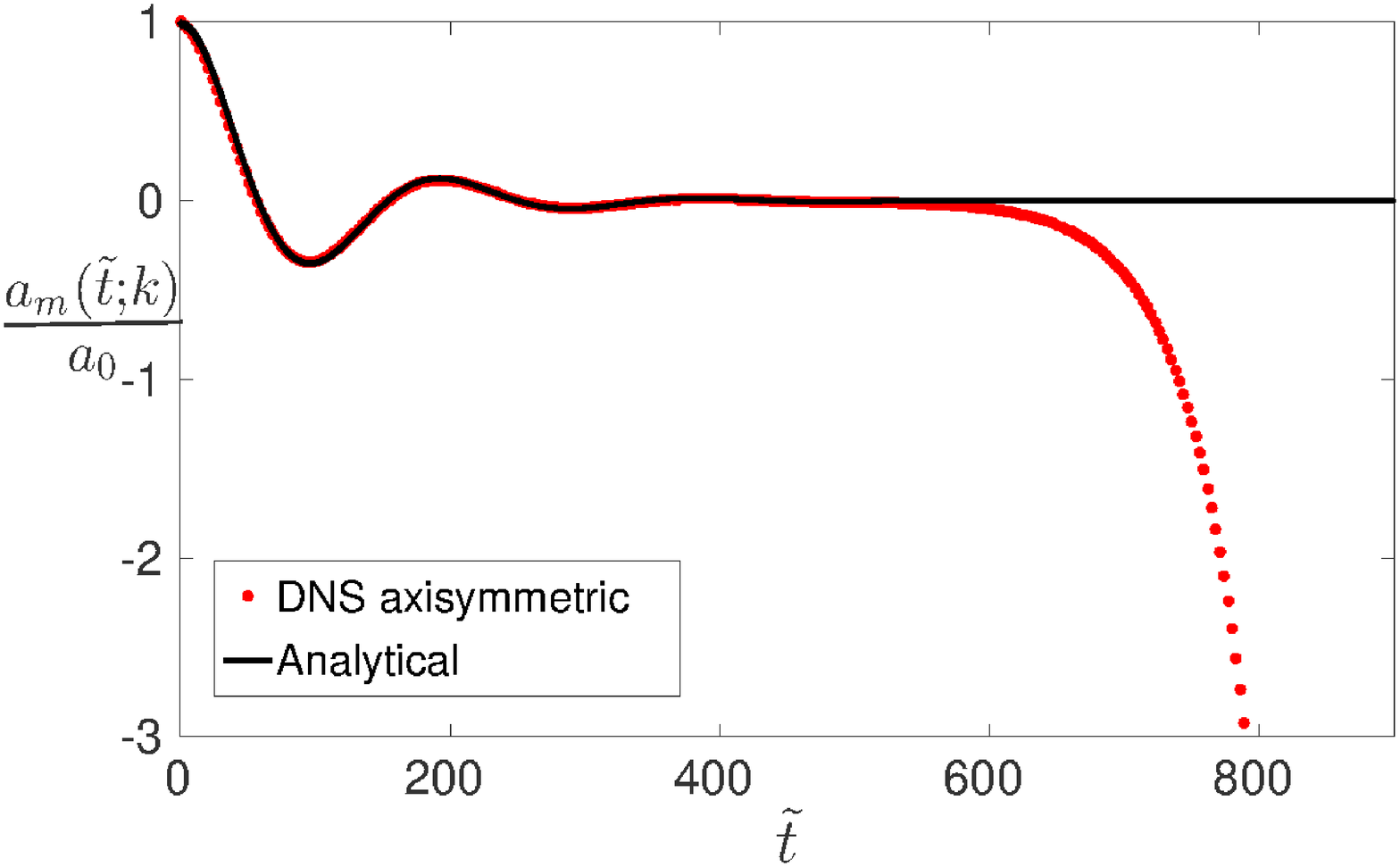}\label{fig8a}}\hspace{4mm}
	\subfloat[Time signal]{%
		\includegraphics[scale=0.22]{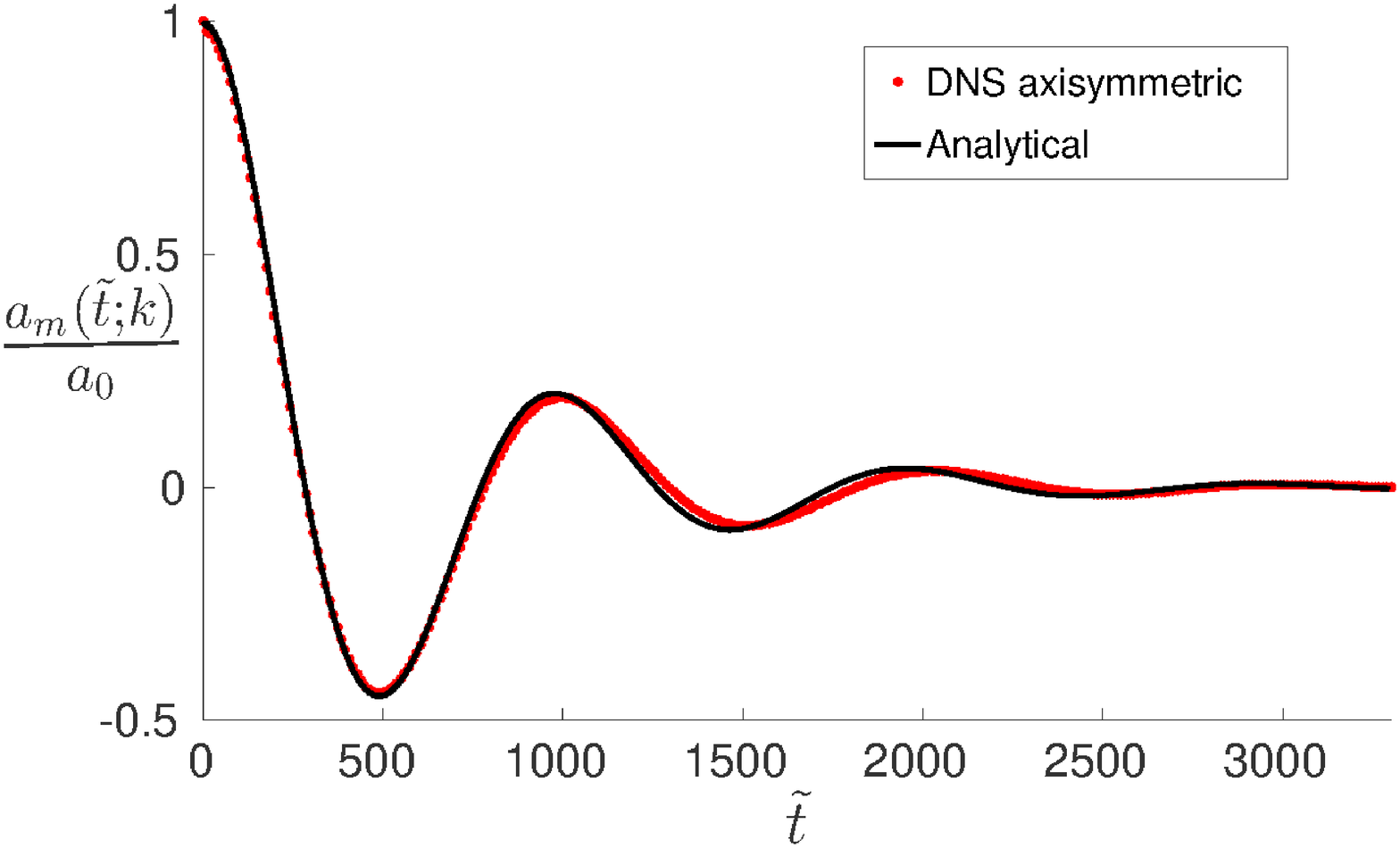}\label{fig8b}}
	\caption{\textbf{Panel a)} Effect of turning-off forcing on RP mode stabilisation. This is the same mode as figure \ref{fig7} with forcing turned off at $\tilde{t} = 485 \approx$ for DNS. Subsequently the RP unstable mode displays unbounded growth. \textbf{Panel b)} Case $2$ in table \ref{tabparam2}: DNS time signal for the mode ($k_0=3.48,m=0$). Stabilisation is seen upto $3000$ forcing cycles with excellent agreement between DNS (axisymmetric) and the solution to equation \ref{eqnIVP7}. Refer stability chart in figure \ref{fig5} for this case with frequency increased to $\Omega=2200$ compared to case $1$.}
	\label{fig8}			
\end{figure}
\subsection{Stablisation of RP modes: DNS results and comparison with theory} 
Figure \ref{fig7} shows stabilisation of the RP mode  $k_0=4.8,m=0$ in DNS, both axisymmetric as well as three dimensional (refer figure \ref{fig4} for stability chart for this case). This is case $1$ in table \ref{tabparam2} and shows stabilisation of the mode $k_0=4.8,m_0=0$ (subscripts $0$ are used for primary modes viz. the modes excited initially in DNS). The solid lines in red and blue are from DNS and nearly overlap. These indicate the amplitude of the interface as a function of time (the interface is tracked at the centre of the domain at $\theta=0$, see figure \ref{fig6}). The signals show stable, underdamped behaviour, decaying to zero after a few hundred forcing cycles ($\approx 400$ cycles). Note the excellent agreement between the DNS signals and the numerical solution to equation \ref{eqnIVP7} indicated by the solid black line. The inset to the figure shows that superposed on the long time underdamped oscillations, are fine scale oscillations arising from the high frequency (compared to the growth rate of the RP mode) forcing imposed on the cylinder. Also shown in figure \ref{fig7} are two more DNS signals, one with forcing $h > h_{\text{cr}1}$ and another with $h < h_{\text{cr}2}$. Both forcing levels are outside the optimum window $h_{\text{cr}1} < h < h_{\text{cr}2}$ and thus stabilisation is not achieved (see figure \ref{fig4} for the optimum forcing window). 

In figure \ref{fig8a}, we further validate the stabilisation obtained in figure \ref{fig7}, by turning off forcing at $\tilde{t}=485$ in DNS. It is seen that the interface destabilises in the absence of forcing indicating that forcing is crucial to the observed stabilisation. 
\begin{figure}
	\centering
	\subfloat[m=0]{%
		\includegraphics[scale=0.25]{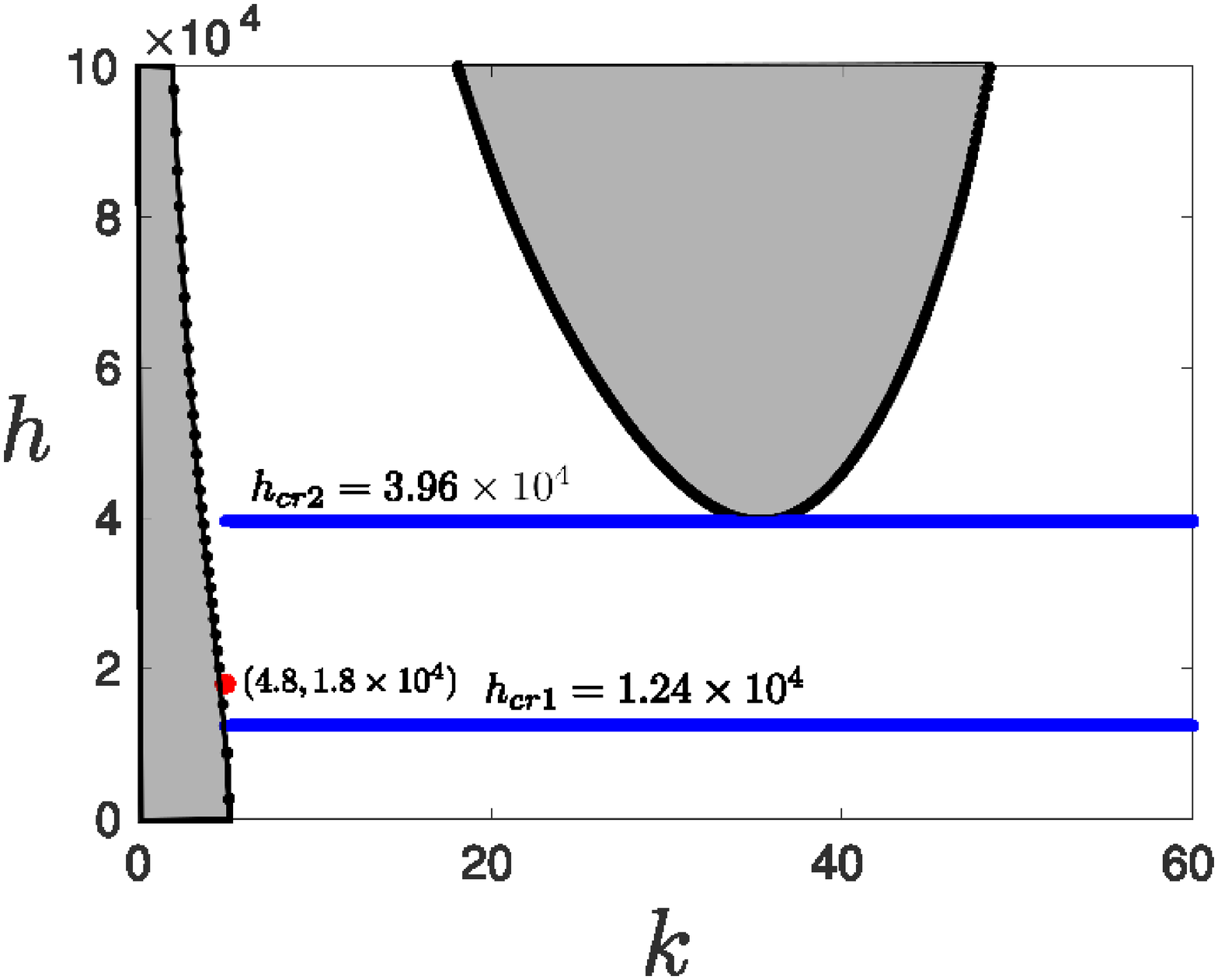}\label{fig9a}}\hspace{4mm}
	\subfloat[Time signal]{%
		\includegraphics[scale=0.24]{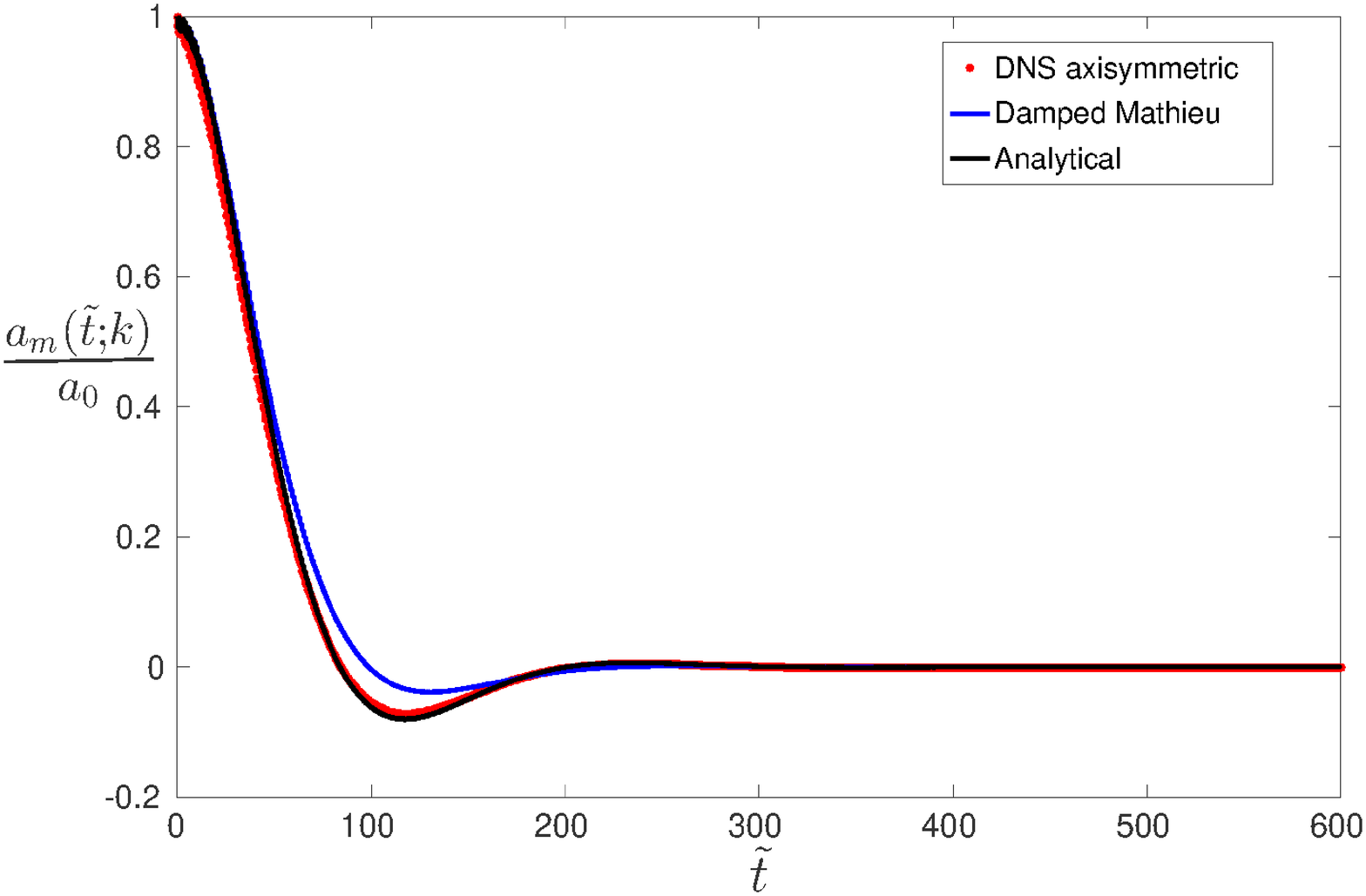}\label{fig9b}}
	\caption{\textbf{Upper panel a)} Stability diagram for case $3$ in table \ref{tabparam2}. The viscosity has been doubled for this case compared to case $1$ in table \ref{tabparam2}. The RP mode $k_0=4.8,m_0=0$ and moderately higher modes are stabilised if $h_{\text{cr}1} < h < h_{\text{cr2}}$. Here $h_{\text{cr1}}=1.24\times 10^4$, and $h_{\text{cr2}}=3.74\times 10^4$ is determined from the non-axisymmetric stability plot for $m=4$ (not shown here). We choose $h=1.8\times10^4$ for stabilisation as indicated by the red dot. \textbf{Lower panel b)} Time signal from axisymmetric DNS showing stabilisation for the RP unstable mode indicated by a red dot in the upper panel viz. $k=4.8,m=0$. Note the overdamped response and the excellent agreement with the soln. to equation \ref{eqnIVP7}. (Blue line) Solution to the damped Mathieu equation equation \ref{eqnDM2}. The analytical response is the solution to equation \ref{eqnDM1}}
	\label{fig9}		
\end{figure}
In figure \ref{fig8b}, we show stabilisation of the RP unstable mode $k_0=3.48,m_0=0$ (Case $2$ in table \ref{tabparam2}). Recall from our discussion in the previous section that the frequency of forcing $\Omega$ was increased to $2200\pi$ for this case, in order to satisfy the ordering $h_{\text{cr}1} < h < h_{\text{cr}2}$ (refer figure \ref{fig5} for stability chart for this case). The figure shows that stabilisation is acheived and sustained for more than $3000$ forcing cycles when the perturbation decays to zero in an underdamped manner. 
\subsection{Damping and the memory term}
We return in this section to a discussion of terms in equation \ref{eqnIVP7} that appear due to viscosity viz. the damping and the memory terms. These terms are physically easiest to intepret in the axisymmetric limit. It is shown in the supplementary material that in this limit, equation \ref{eqnIVP7} reduces to
  \begin{eqnarray}
 	&&\frac{d^2a_0}{dt^2} + 2\nu k^2\left(1+ \frac{\mathrm{I}_0''(kR_0)}{\mathrm{I}_0(kR_0)}\right)\frac{da_0}{dt}   + \frac{\mathrm{I}_0'(kR_0)}{\mathrm{I}_0(kR_0)}\left[ \frac{T}{\rho R_0^3}kR_0 (k^2R_0^2-1)+ kh \cos(\Omega t) \right] a_0(t) \nonumber\\&& +\frac{4\nu^2 k^4}{\mathrm{I}_0(kR_0)}\int_{0}^{t}\hat{\textbf{L}}^{-1}\left[\mathcal{K}(s)\right]\frac{da_0}{dt}(t - \tau) d\tau = 0\label{eqnDM1}\\
 	&&\text{where}, \quad \mathcal{K}(s) = \left( \frac{\mathrm{I}_0''(kR_0)}{s} -\frac{l}{k}\frac{\mathrm{I}_0'(kR_0)\mi_0^{''}(lR_0)}{s\mi_0^{'}(lR_0)}\right) \nonumber
 \end{eqnarray}
If we temporarily disregard the memory term in equation \ref{eqnDM1}, then it is clear that the rest of equation constitutes a damped Mathieu equation i.e. the damped version of equation \ref{eqn3} for $m=0$ (axisymmetric). This is 
  \begin{eqnarray}
  &&\frac{d^2a_0}{dt^2} + 2\nu k^2\left(1+ \frac{\mathrm{I}_0''(kR_0)}{\mathrm{I}_0(kR_0)}\right)\frac{da_0}{dt}   + \frac{\mathrm{I}_0'(kR_0)}{\mathrm{I}_0(kR_0)}\left[ \frac{T}{\rho R_0^3}kR_0 (k^2R_0^2-1)+ kh \cos(\Omega t) \right] a_0(t) = 0 \nonumber \\ \label{eqnDM2}
  \end{eqnarray}
Equation \ref{eqnDM2} is the cylindrical analogue of its Cartesian counterpart which has been discussed in \cite{kumar1994parametric,cerda1998faraday} for viscous Faraday waves over a flat interface (see equation 4.21 in \cite{kumar1994parametric} or equation 3.4 in \cite{cerda1998faraday}). In order to put this analogy on a sound footing, we take the limit $R_0\rightarrow \infty$ (for fixed $k$) on equation \ref{eqnDM2} expecting to recover results relevant to a flat interface (as $R_0\rightarrow\infty$, the cylinder locally becomes flat). Using the identity $\lim_{z\rightarrow \infty}\mi_{0}^{''}(z)/\mi_{0}(z) = 1$, it is seen that the coefficient of the second term in \ref{eqnDM2} in this limit, reduces to the damping coefficient of viscous capillary waves (deep water) on a flat interface viz. $4\nu k^2$, which is the same as estimated in \cite{kumar1994parametric,cerda1998faraday}. Note that the damping factor $4\nu k^2$ for a flat interface is obtained by estimating disspation for potential flow \citep{kumar1994parametric}. By analogy it may similarly be expected that the pre-factor $2\nu k^2\left(1+ \frac{\mathrm{I}_0''(kR_0)}{\mathrm{I}_0(kR_0)}\right)$ in equation \ref{eqnDM2} arises from the damping of \textit{potential} flow \citep{patankar2018faraday} in the liquid cylinder. It has been verified that this is correct and the factor $2\nu k^2\left(1+ \frac{\mathrm{I}_0''(kR_0)}{\mathrm{I}_0(kR_0)}\right)$ indeed agrees with the damping predicted by the dispersion relation in equation 5.10 of \cite{wang2005pressure} which was obtained through a \textit{viscous potential flow calculation} (VCVPF in their terminology with a crucial viscous pressure correction)

Turning now to the memory term in equation \ref{eqnDM1}, we note that it does not depend on the forcing strength $h$. Thus it persists even in the unforced limit ($h\rightarrow0$), in which case equation \ref{eqnDM1} becomes one governing free perturbations. This equation was derived earlier by \cite{berger1988initial} by solving the corresponding IVP with $h=0$ and we have verified that the unforced limit of equation \ref{eqnDM1} agrees with the equation of \cite{berger1988initial} (see supplementary material). The Laplace inversion of $\mathcal{K}(s)$ in equation \ref{eqnDM1} is analytically feasible and maybe expressed as infinite summation over integrals from residue theory (see expression 79 in \cite{berger1988initial}). For convenience, we reproduce this here as the term on the right hand side of equation \ref{eqnDM3} (the damping term in equation \ref{eqnDM3} has been slightly modified from \cite{berger1988initial} but is exactly equivalent to his expression)
\begin{eqnarray}
	&& \frac{d^2a_0}{dt^2} + 2\nu k^2\left(1+ \frac{\mathrm{I}_0''(kR_0)}{\mathrm{I}_0(kR_0)}\right)\frac{da_0}{dt}(t) + \left[\frac{T}{\rho R_0^3} kR_0 \left(k^2R_0^2-1\right)\frac{\mi_1(kR_0)}{\mi_0(kR_0)}\right]a_0(t) \nonumber \\
	&=& \frac{8\nu^2k^3}{R_0}\frac{\mi_0(kR_0)}{\mi_1(kR_0)}\int_{0}^{t}\frac{da_0(t')}{dt^{'}}\exp\left(-\nu k^2(t-t^{'})\right)\displaystyle\sum_{j_n}\frac{\exp\left[-\left(\frac{\nu}{R_0^2}\right)j_n^2(t-t^{'})\right]}{1 + \left(\frac{R_0k}{j_n}\right)^2}
	\label{eqnDM3}
\end{eqnarray}
where $j_n$ represents the $n$th (non-zero) zero of $J_1(j_n)=0$ \citep{berger1988initial}. The origin of the infinite summation in \ref{eqnDM3} may be rationalised as follows: the initial condition of zero vorticity and surface deformation (i.e. $\eta(z,\theta,0)=a_0\cos(k_0z)$) excites all modes in the spectrum (viz. two capillary modes and a countable infinite set of hydrodynamic modes \citep{garcia2008normal}). The excitation of the countably infinite set of hydrodynamic modes (which are all purely damped modes) produces the infinite summation in the analytical expression for $a_0(t;k)$ also manifesting as the memory term(s) in equation \ref{eqnDM3}. These conclusions for free perturbations on a cylinder have analogues on a flat surface (e.g. see equation 2.30 in \cite{cerda1998faraday} which expresses the amplitude as a sum over two capillary modes and an infinite sum over the hydrodynamic modes). 

Physically, the presence of the memory term implies that the damping seen in DNS contains contributions not only from the potential part of the flow (as is modelled correctly by the damped Mathieu equation equation \ref{eqnDM2}) but also from the memory term(s) which arise due to the boundary layer at the free surface. We find that the contribution of the memory term in equation \ref{eqnDM2} increases as the kinematic viscosity of the fluid is increased and is the largest (in the axisymmetric limit being studied here), when viscosity is sufficiently large for the stabilised response of the liquid cylinder to be overdamped. Figure \ref{fig9b} depicts this for the RP mode $k_0=4.8,m_0=0$ (Case $3$ in table \ref{tabparam2}) highlighting the difference between the solution to the damped Mathieu equation \ref{eqnDM2} and the integro-differential equation \ref{eqnDM1}. It is seen that at intermediate time ($80 < \tilde{t} < 100$), the damped Mathieu equation \ref{eqnDM2}, underpredicts the damping that is seen in the DNS and in equation \ref{eqnDM1}. The corresponding stability chart with the optimal level of forcing for stabilisation is indicated in the upper panel of figure \ref{fig9a}.

We conclude this study with a discussion on the limitation of the present stabilisation technique viz. that it does not stabilize the entire RP unstable spectrum at any finite level of forcing, but only modes with $k > k_0$. This arises from the infinitely long cylinder assumption that we have made allowing all modes from $0 < k_0 < \infty$ to be present. In practise we expect to encounter liquid cylinders of finite length typically confined between supports. The boundary conditions at the end-points (e.g. pinned, see \cite{sanz_1985}) can substantially modify the nature of the eigenmodes in the $z$ direction compared to the Fourier modes that we have assumed here. As remarked in the introduction, stabilisation of capillary-bridges is an active area of research and the specific problem of dynamic stabilisation of a liquid bridge is under investigation and will be reported in future.

\section{Conclusions}
In this study, we have proposed dynamic stabilisation of RP unstable modes on a viscous liquid cylinder subject to radial, harmonic forcing. We use linearised, viscous stability analysis employing the toroidal-poloidal decomposition \citep{marques1990boundary,boronski2007poloidal}. It is demonstrated that for a viscous fluid, by suitably tuning the frequency of forcing and optimally choosing its strength, not only can a chosen axisymmetric RP mode ($k_0$) be stabilised but also all moderately large integral multiples of $k_0$, both axisymmetric and three-dimensional, can be prevented from destabilising the cylinder. Direct numerical simulations have been used to validate theoretical predictions demonstrating stabilisation upto hundreds of forcing cycles, in marked contrast to our earlier inviscid study \citep{patankar2018faraday} where stabilisation could not be achieved. We have shown that viscosity plays a crucial role in this as it enables the upper critical threshold of forcing to be greater than zero $h_{\text{cr}2} > 0$, unlike the inviscid case. It is demonstrated that one can tune the forcing frequency $\Omega$ such that the optimal strength of forcing satisfies satisfy $h_{\text{cr}1} < h < h_{\text{cr}2}$. 

Additionally, we have also solved the initial-value problem (IVP) corresponding to surface deformation and zero vorticity initial conditions, leading to a novel integro-differential equation governing the (linearised) amplitude of three-dimensional Fourier modes on the cylinder. This equation is non-local in time and represents the cylindrical analogue of the one governing Faraday waves on a flat interface \citep{beyer1995faraday,cerda1997faraday}. Our equation generalises to the viscous case the Mathieu equation that was derived in \cite{patankar2018faraday}. In the axisymmetric limit, we have proven that the memory term in the equation is inherited from the unforced problem and represents the excitation of damped hydrodynamic modes. We find that the contribution from this term is the highest when fluid viscosity is taken to be sufficiently large such that the stabilised response of the RP mode is overdamped.
The stabilisation strategy that has been proposed here can in-principle be used to stabilise any axisymmetric RP mode of wavenumber $k_0$. In practise, as $k_0$ gets smaller (longer modes), the threshold frequency increases sharply and compressibility effects can become important. We have also seen that modes which satisfy $ k < k_0$, are still unstable although they are inaccessible to our numerical simulations due to the periodic nature of the boundary conditions. This is proposed for future study wherein we will investigate dynamic stabilisation of liquid bridges held between substrates as well as stabilisation of thin films coating a hollow tube pulsating radially in time. The latter situation also offers a way to practically realize the radial, oscillating body force which has been applied here.

We conclude with an interesting analogy of the present study with that of \cite{woods1995instability}. In our study, there is a range of long waves ($ k < R_0^{-1}$) which are linearly unstable when there is no forcing ($h=0$). For fixed viscosity of the liquid and through optimal choice of the strength ($h$) and frequency of forcing ($\Omega$), we have demonstrated stabilisation of these hitherto unstable RP modes. A nearly analogous situation arises in flow over an infinitely long inclined plane where the base-flow is linearly unstable to long gravity waves \citep{yih1967instability,benjamin1954stability} and may be stabilised by subjecting the plane to vertical oscillation. Fig. $4$ of the study by \cite{woods1995instability}, bears a strong qualitative resemblance to our axisymmetric stability charts (inset of figure \ref{fig4a}).
\section*{Acknowledgements}
We acknowledge support from DST-SERB vide grants \\
\#EMR/2016/000830, \#MTR/2019/001240 and \#CRG/2020/003707 and an IRCC-IITB startup grant to RD. The Ph.D. fellowship for SP is supported through grants from DST-SERB (\#EMR/2016/000830) and IRCC-IITB. SB acknowledges fellowship support through the Prime Minister's Research Fellowship (PMRF), Govt. of India. We thank Dr. Palas Kumar Farsoiya for helpful discussions and assistance in the early stage of this study.

\section*{Appendix A: expressions for coefficients}
Expressions for $\mathcal{A}(s), \mathcal{C}(s)$ and $\mathcal{D}(s)$ used in solution to the IVP are provided below:
\begin{eqnarray}
&&\mathcal{A}(s) = 2k^2l\mathrm{I}_m'(lR_0)\mathrm{I}_m'(kR_0) \left\{ \frac{\left( l^2 + k^2 \right)\Lambda_3 -  2k^2\Lambda_2(s)}{\beta(s)} \right\} \left[s\tilde{a}_m - a_0\right] \label{25}\\
&&\mathcal{C}(s) = \frac{2mk^3}{R_0}\mathrm{I}_m(lR_0)\mathrm{I}_m'(kR_0)\left(\frac{\Lambda_1(s) - \Lambda_3 }{\beta(s)}\right)\left[s\tilde{a}_m - a_0\right]\\
&&\mathcal{D}(s) = \frac{ml}{R_0}\mathrm{I}_m(lR_0)\mathrm{I}_m'(lR_0)\left\{\frac{ 2k^2\Lambda_2(s)-\left( l^2 + k^2 \right)\Lambda_1(s) }{\beta(s)}\right\}\left[s\tilde{a}_m(s;k) - a_0\right]\\
&& \textrm{where} \quad \beta(s) \equiv \textrm{Det}
\begin{bmatrix}
\frac{m}{R_0}\mathrm{I}_m(lR_0) && kl\mathrm{I}_m'(lR_0) && k^2\mathrm{I}_m'(kR_0)\\
\frac{mk}{R_0}\mathrm{I}_m(lR_0) && \left( l^2 + k^2\right)l\mathrm{I}_m'(lR_0) && 2k^3\mathrm{I}_m'(kR_0)\\
\frac{m}{R_0}\mathrm{I}_m(lR_0)\Lambda_1(s) && 2kl\mathrm{I}_m'(lR_0)\Lambda_2(s) && 2k^2\mathrm{I}_m'(kR_0)\Lambda_3
\end{bmatrix} \nonumber\\
&& \hspace{2cm} = \frac{mlk^2}{R_0}\mathrm{I}_m(lR_0)\mathrm{I}_m'(lR_0)\mathrm{I}_m'(kR_0)\Lambda(s),\\
&&l^2 \equiv k^2 + \frac{s}{\nu},\quad \Lambda(s) \equiv \left(k^2 - l^2\right)\Lambda_1(s) - 2k^2\Lambda_2(s) + 2l^2\Lambda_3, \\
&&\Lambda_1(s) \equiv 1 - \frac{lR_0}{m^2}\frac{\mathrm{I}_m'(lR_0)}{\mathrm{I}_m(lR_0)} + \frac{R_0^2l^2}{m^2}\frac{\mathrm{I}_m''(lR_0)}{\mathrm{I}_m(lR_0)},\\
&&\Lambda_{2}(s) \equiv 1 - \frac{1}{lR_0}\frac{\mathrm{I}_m(lR_0)}{\mathrm{I}_m'(lR_0)}\quad \textrm{and}\; \Lambda_3 = 1 - \frac{1}{kR_0}\frac{\mathrm{I}_m(kR_0)}{\mathrm{I}_m'(kR_0)}. 
\end{eqnarray}
\section*{Appendix B}
For axisymmetric perturbation $m = 0$, the equation governing $a_0(t;k)$ may be written in the time domain as (see supplementary material)
\begin{eqnarray}
&&\frac{d^2a_0}{dt^2} + 2\nu k^2\left(1+ \frac{\mathrm{I}_0''(kR_0)}{\mathrm{I}_0(kR_0)}\right)\frac{da_0}{dt}   + \frac{\mathrm{I}_0'(kR_0)}{\mathrm{I}_0(kR_0)}\left[  \frac{T}{\rho R_0^3} kR_0\left(k^2R_0^2-1\right) + kh\cos(\Omega t) \right] a_0(t) \nonumber\\&& +\frac{4\nu^2 k^4}{\mathrm{I}_0(kR_0)}\int_{0}^{t}\hat{\textbf{L}}^{-1}\left[\mathcal{K}(s)\right]\frac{da_0}{dt}(t - \tau) d\tau = 0 \label{APB1}\\
&&\text{where}, \quad \mathcal{K}(s) = \left( \frac{\mathrm{I}_0''(kR_0)}{s} -\frac{l}{k}\frac{\mathrm{I}_0'(kR_0)\mi_0^{''}(lR_0)}{s\mi_0^{'}(lR_0)}\right) \nonumber
\end{eqnarray}
Using the identity $ \mi_0^{'}(kR_0) = \mi_1(kR_0)$ and $ \mi_1^{'}(kR_0) = \left( \mi_0(kR_0) - \frac{1}{kR_0}\mi_1(kR_0)\right)$, we obtain 
\begin{eqnarray}
&&\frac{d^2a_0}{dt^2}  + 4\nu k^2\left\{1 - \frac{1}{2k R_0}\cdot\frac{\mathrm{I}_1(kR_0)}{\mathrm{I}_0(kR_0)}\right\}\frac{da_0}{dt} + \frac{\mathrm{I}_1(kR_0)}{\mathrm{I}_0(kR_0)}\left[\frac{T}{\rho R_0^3} kR_0\left(k^2R_0^2-1\right) + hk \cos\left(\Omega t \right) \right]a_0(t) \nonumber\\ 
&& + 4\nu^2k^4\int_{0}^{t}\mathcal{K}(\tau)\frac{da_0}{dt}(t - \tau)d\tau = 0 \label{APB2}\\
\textrm{where,} && \Tilde{\mathcal{K}}(s) = {\hat{\textbf{L}}}\left[\mathcal{K}(\tau)\right] =  \frac{1}{s}\left\{ 1 - \frac{l}{k}\cdot\frac{\mathrm{I}_1(kR_0)}{\mathrm{I}_0(kR_0)}\cdot \frac{\mathrm{I}_0(lR_0)}{\mathrm{I}_1(lR_0)}\right\}\nonumber
\end{eqnarray}
In the limit, $R_0 \rightarrow \infty$, equation \ref{APB2} becomes
\begin{eqnarray}
&&\frac{d^2a_0}{dt^2} + 4\nu k^2 \frac{da_0}{dt} + \left[\frac{Tk^3}{\rho} + hk \cos\left(\Omega t \right) \right]a(t) + 4\nu^2 k^4\int_{-\infty}^{t}\mathcal{K}^{(\infty)}(t - \tau)\frac{da_0}{d\tau}(\tau) d\tau = 0\nonumber\\\label{APB3}\\
&& \textrm{where}\quad \tilde{\mathcal{K}}^{(\infty)}(s) = {\hat{\textbf{L}}}\left[\mathcal{K}^{(\infty)}(t)\right] = \frac{1}{s}\left\{1 - \frac{l}{k}\right\} = \frac{1}{s} - \frac{1}{k\sqrt{\nu}} \cdot \frac{\sqrt{s + \nu k^2}}{s}\nonumber
\end{eqnarray}
From \cite{erdelyi1954tables}, we can analytically invert $\tilde{\mathcal{K}}^{(\infty)}(s)$ to write 
\begin{eqnarray}
\mathcal{K}^{(\infty)}(t) &=& 1 - \frac{1}{k\sqrt{\nu}}\left[\frac{1}{\sqrt{\pi t}}e^{-\nu k^2 t} + k\sqrt{\nu} \cdot \frac{1}{\sqrt{\pi}}\int_{0}^{\nu k^2t}\frac{e^{-t'}}{\sqrt{t'}} dt'   \right]\nonumber\\
\textrm{or,}\quad \mathcal{K}^{(\infty)}(t) &=& 1 - \frac{1}{k\sqrt{\nu \pi}}\cdot \frac{e^{-\nu k^2 t}}{\sqrt{t}} - \frac{1}{\sqrt{\pi}}\int_{0}^{\nu k^2t}\frac{e^{-t'}}{\sqrt{t'}} dt'\nonumber\\
\textrm{or,}\quad \mathcal{K}^{(\infty)}(t - \tau) &=& 1 - \frac{1}{k\sqrt{\nu \pi}}\cdot \frac{e^{-\nu k^2(t - \tau)}}{\sqrt{t - \tau}} - \frac{1}{\sqrt{\pi}}\int_{0}^{\nu k^2(t - \tau)}\frac{e^{-t'}}{\sqrt{t'}} dt' \label{APB4}
\end{eqnarray}
Substituting expression \ref{APB4} in equation \ref{APB3},
\begin{eqnarray}
&&\frac{d^2a_0}{dt^2} + 4\nu k^2 \frac{da_0}{dt} + 4\nu^2k^4 a_0(t) + \left[\frac{Tk^3}{\rho} + hk \cos\left(\Omega t \right) \right]a_0(t) - \frac{4\nu^{3/2}k^3}{\sqrt{\pi}}\int_{-\infty}^{t}\frac{e^{-\nu k^2(t - \tau)}}{\sqrt{t - \tau}}\frac{da_0}{d\tau}(\tau)d\tau \nonumber \\ 
&&- \frac{4\nu^2k^4}{\sqrt{\pi}}\int_{-\infty}^{t}\Phi(t - \tau)\frac{da_0}{d\tau}(\tau) d\tau = 0 \label{APB5}\\
&&\textrm{where} \quad \Phi(t - \tau) = \int_{0}^{\nu k^2(t - \tau)}\frac{e^{-t'}}{\sqrt{t'}} dt' \nonumber
\end{eqnarray}
Integrating by parts the last integral term of above equation and using the shorthand notation $\frac{d}{dt} \equiv \partial_t$,  
\begin{eqnarray}
&&\frac{1}{k}{\left( \partial_t + 2\nu k^2 \right)}^2 a_0(t) + \left[ \frac{Tk^2}{\rho} + h\cos\left(\Omega t \right) \right]a(t) - \frac{4\nu^{3/2}k^2}{\sqrt{\pi}}\int_{-\infty}^{t}\frac{e^{-\nu k^2(t - \tau)}}{\sqrt{t - \tau}}\partial_{\tau}a_0(\tau)d\tau \nonumber\\
&&- \frac{4\nu^2k^3}{\sqrt{\pi}}\left[\Phi(t - \tau)a_0(\tau) \Bigg|_{\tau  = -\infty}^{\tau = t} + k\sqrt{\nu}\int_{-\infty}^{t}\frac{e^{-\nu k^2(t - \tau)}}{\sqrt{t - \tau}}a_0(\tau)d\tau  \right] = 0\nonumber\\
\textrm{or,}&&\frac{1}{k}{\left( \partial_t + 2\nu k^2 \right)}^2 a_0(t) + \left[ \frac{Tk^2}{\rho} + h\cos\left(\Omega t \right) \right]a_0(t) - \frac{4\nu^{3/2}k^2}{\sqrt{\pi}}\int_{-\infty}^{t}\frac{e^{-\nu k^2(t - \tau)}}{\sqrt{t - \tau}}\partial_{\tau}a_0(\tau)d\tau \nonumber \\
&& - \frac{4\nu^{5/2}k^4}{\sqrt{\pi}}\int_{-\infty}^{t}\frac{e^{-\nu k^2(t - \tau)}}{\sqrt{t - \tau}}a_0(\tau)d\tau = 0 \nonumber\\
\textrm{or,}&&\frac{1}{k}{\left( \partial_t + 2\nu k^2 \right)}^2 a_0(t) + \left[ \frac{Tk^2}{\rho} + h\cos\left(\Omega t \right) \right]a_0(t)\nonumber \\
&& - 2\nu k^2\frac{2\sqrt{\nu}}{\pi}\int_{-\infty}^{t}\mathrm{G}(t - \tau)e^{-\nu k^2(t - \tau)}\left(\partial_\tau + \nu k^2\right)a_0(\tau)d\tau = 0 \label{APB6} \\
&&\text{where} \quad \mathrm{G}(t - \tau) \equiv \sqrt{\frac{\pi}{t - \tau}} \nonumber
\end{eqnarray}
Equation \ref{APB6}  matches with equation 44 in \cite{beyer1995faraday} in the deep water limit.

\bibliographystyle{jfm}
\bibliography{jfm-instructions}

\begin{thebibliography}{88}
\expandafter\ifx\csname natexlab\endcsname\relax\def\natexlab#1{#1}\fi
\def\au#1{#1} \def\ed#1{#1} \def\yr#1{#1}\def\at#1{#1}\def\jt#1{\textit{#1}}
  \def\bt#1{#1}\def\bvol#1{\textbf{#1}} \def\vol#1{#1} \def\pg#1{#1}
  \def\publ#1{#1}\def\arxiv#1{#1}\def\org#1{#1}\def\st#1{\textit{#1}}

\bibitem[Adou \& Tuckerman(2016)]{adou2016faraday}
{\sc \au{Adou, Ali-higo~Ebo} \& \au{Tuckerman, Laurette~S}} \yr{2016}
  \at{Faraday instability on a sphere: Floquet analysis}.  \jt{Journal of Fluid
  Mechanics}  \bvol{805},  \pg{591--610}.

\bibitem[F.~W. J. Olver~et. al.(2021)]{NIST:DLMF}
{\sc \au{F.~W. J. Olver~et. al., eds.}} \yr{2021} Nist digital library of
  mathematical functions. \url{http://dlmf.nist.gov/, Release 1.1.3 of
  2021-09-15}.

\bibitem[Arbell \& Fineberg(2000)]{arbell2000temporally}
{\sc \au{Arbell, H} \& \au{Fineberg, J}} \yr{2000}  \at{Temporally harmonic
  oscillons in newtonian fluids}.  \jt{Physical Review Letters}  \bvol{85}~(4),
   \pg{756}.

\bibitem[Basak {\em et~al.\/}(2021)Basak, Farsoiya \&
  Dasgupta]{basak_farsoiya_dasgupta_2021}
{\sc \au{Basak, Saswata}, \au{Farsoiya, Palas~Kumar} \& \au{Dasgupta, Ratul}}
  \yr{2021}  \at{Jetting in finite-amplitude, free, capillary-gravity waves}.
  \jt{Journal of Fluid Mechanics}  \bvol{909},  \pg{A3}.

\bibitem[Batson {\em et~al.\/}(2013)Batson, Zoueshtiagh \&
  Narayanan]{batson_zoueshtiagh_narayanan_2013}
{\sc \au{Batson, W.}, \au{Zoueshtiagh, F.} \& \au{Narayanan, R.}} \yr{2013}
  \at{The faraday threshold in small cylinders and the sidewall non-ideality}.
  \jt{Journal of Fluid Mechanics}  \bvol{729},  \pg{496–523}.

\bibitem[Bechhoefer {\em et~al.\/}(1995)Bechhoefer, Ego, Manneville \&
  Johnson]{bechhoefer_ego_manneville_johnson_1995}
{\sc \au{Bechhoefer, John}, \au{Ego, Valerie}, \au{Manneville, Sebastien} \&
  \au{Johnson, Brad}} \yr{1995}  \at{An experimental study of the onset of
  parametrically pumped surface waves in viscous fluids}.  \jt{Journal of Fluid
  Mechanics}  \bvol{288},  \pg{325–350}.

\bibitem[Benilov(2016)]{benilov2016stability}
{\sc \au{Benilov, ES}} \yr{2016}  \at{Stability of a liquid bridge under
  vibration}.  \jt{Physical Review E}  \bvol{93}~(6),  \pg{063118}.

\bibitem[Benjamin \& Ursell(1954)]{benjamin1954stability}
{\sc \au{Benjamin, Thomas~Brooke} \& \au{Ursell, Fritz~Joseph}} \yr{1954}
  \at{The stability of the plane free surface of a liquid in vertical periodic
  motion}.  \jt{Proceedings of the Royal Society of London. Series A.
  Mathematical and Physical Sciences}  \bvol{225}~(1163),  \pg{505--515}.

\bibitem[Berger(1988)]{berger1988initial}
{\sc \au{Berger, SA}} \yr{1988}  \at{Initial-value stability analysis of a
  liquid jet}.  \jt{SIAM Journal on Applied Mathematics}  \bvol{48}~(5),
  \pg{973--991}.

\bibitem[Beyer \& Friedrich(1995)]{beyer1995faraday}
{\sc \au{Beyer, J} \& \au{Friedrich, R}} \yr{1995}  \at{Faraday instability:
  linear analysis for viscous fluids}.  \jt{Physical Review E}  \bvol{51}~(2),
  \pg{1162}.

\bibitem[Binz {\em et~al.\/}(2014)Binz, Rohlfs \& Kneer]{binz2014direct}
{\sc \au{Binz, Matthias}, \au{Rohlfs, Wilko} \& \au{Kneer, Reinhold}} \yr{2014}
   \at{Direct numerical simulations of a thin liquid film coating an axially
  oscillating cylindrical surface}.  \jt{Fluid Dynamics Research}
  \bvol{46}~(4),  \pg{041402}.

\bibitem[Boffetta {\em et~al.\/}(2019)Boffetta, Magnani \&
  Musacchio]{boffetta2019suppression}
{\sc \au{Boffetta, Guido}, \au{Magnani, Marta} \& \au{Musacchio, Stefano}}
  \yr{2019}  \at{Suppression of rayleigh-taylor turbulence by time-periodic
  acceleration}.  \jt{Physical Review E}  \bvol{99}~(3),  \pg{033110}.

\bibitem[Boronski \& Tuckerman(2007)]{boronski2007poloidal}
{\sc \au{Boronski, Piotr} \& \au{Tuckerman, Laurette~S}} \yr{2007}
  \at{Poloidal--toroidal decomposition in a finite cylinder. i: Influence
  matrices for the magnetohydrodynamic equations}.  \jt{Journal of
  Computational Physics}  \bvol{227}~(2),  \pg{1523--1543}.

\bibitem[Cerda \& Tirapegui(1997)]{cerda1997faraday}
{\sc \au{Cerda, Enrique} \& \au{Tirapegui, Enrique}} \yr{1997}  \at{Faraday's
  instability for viscous fluids}.  \jt{Physical review letters}
  \bvol{78}~(5),  \pg{859}.

\bibitem[Cerda \& Tirapegui(1998)]{cerda1998faraday}
{\sc \au{Cerda, EA} \& \au{Tirapegui, EL}} \yr{1998}  \at{Faraday's instability
  in viscous fluid}.  \jt{Journal of Fluid Mechanics}  \bvol{368},
  \pg{195--228}.

\bibitem[Chandrasekhar(1981)]{chandrasekhar1981hydrodynamic}
{\sc \au{Chandrasekhar, S}} \yr{1981}  \at{Hydrodynamic and hydromagnetic
  stability} .

\bibitem[Chen \& Tsamopoulos(1993)]{chen1993nonlinear}
{\sc \au{Chen, Tay-Yuan} \& \au{Tsamopoulos, John}} \yr{1993}  \at{Nonlinear
  dynamics of capillary bridges: theory}.  \jt{Journal of Fluid Mechanics}
  \bvol{255},  \pg{373--409}.

\bibitem[Driessen(2013)]{driessen2013drop}
{\sc \au{Driessen, Theo}} \yr{2013}  \at{Drop formation from axi-symmetric
  fluid jets}.  \jt{Diss. University of Twente} .

\bibitem[Driessen {\em et~al.\/}(2014)Driessen, Sleutel, Dijksman, Jeurissen \&
  Lohse]{driessen2014control}
{\sc \au{Driessen, Theo}, \au{Sleutel, Pascal}, \au{Dijksman, Frits},
  \au{Jeurissen, Roger} \& \au{Lohse, Detlef}} \yr{2014}  \at{Control of jet
  breakup by a superposition of two rayleigh-plateau-unstable modes}.
  \jt{Journal of fluid mechanics}  \bvol{749},  \pg{275--296}.

\bibitem[Edwards \& Fauve(1994)]{edwards1994patterns}
{\sc \au{Edwards, W~Stuart} \& \au{Fauve, S}} \yr{1994}  \at{Patterns and
  quasi-patterns in the faraday experiment}.  \jt{Journal of Fluid Mechanics}
  \bvol{278},  \pg{123--148}.

\bibitem[Erdelyi {\em et~al.\/}(1954)Erdelyi, Magnus, Oberhettinger \&
  Tricomi]{erdelyi1954tables}
{\sc \au{Erdelyi, Arthur}, \au{Magnus, Wilhelm}, \au{Oberhettinger, Fritz} \&
  \au{Tricomi, Francesco~G}} \yr{1954} {\em Tables of Integral Transforms:
  Vol.: 2\/}.  \publ{McGraw-Hill Book Company, Incorporated}.

\bibitem[Faraday(1837)]{faraday1837peculiar}
{\sc \au{Faraday, Michael}} \yr{1837} On a peculiar class of acoustical
  figures; and on certain forms assumed by groups of particles upon vibrating
  elastic surfaces.  \bt{In {\em Abstracts of the Papers Printed in the
  Philosophical Transactions of the Royal Society of London\/}},  \pg{pp.
  49--51}. The Royal Society London.

\bibitem[Farsoiya {\em et~al.\/}(2017)Farsoiya, Mayya \&
  Dasgupta]{farsoiya2017axisymmetric}
{\sc \au{Farsoiya, Palas~Kumar}, \au{Mayya, YS} \& \au{Dasgupta, Ratul}}
  \yr{2017}  \at{Axisymmetric viscous interfacial oscillations--theory and
  simulations}.  \jt{Journal of Fluid Mechanics}  \bvol{826},  \pg{797--818}.

\bibitem[Farsoiya {\em et~al.\/}(2021)Farsoiya, Popinet \&
  Deike]{farsoiya2021bubble}
{\sc \au{Farsoiya, Palas~Kumar}, \au{Popinet, St{\'e}phane} \& \au{Deike, Luc}}
  \yr{2021}  \at{Bubble-mediated transfer of dilute gas in turbulence}.
  \jt{Journal of Fluid Mechanics}  \bvol{920}.

\bibitem[Farsoiya {\em et~al.\/}(2020)Farsoiya, Roy \&
  Dasgupta]{farsoiya_roy_dasgupta_2020}
{\sc \au{Farsoiya, Palas~Kumar}, \au{Roy, Anubhab} \& \au{Dasgupta, Ratul}}
  \yr{2020}  \at{Azimuthal capillary waves on a hollow filament – the
  discrete and the continuous spectrum}.  \jt{Journal of Fluid Mechanics}
  \bvol{883},  \pg{A21}.

\bibitem[Fauve(1998)]{fauve1998waves}
{\sc \au{Fauve, S}} \yr{1998}  \at{Waves on interfaces}.  \bt{In {\em Free
  Surface Flows\/}},  \pg{pp. 1--44}.  \publ{Springer}.

\bibitem[Garc{\'\i}a \& Gonz{\'a}lez(2008)]{garcia2008normal}
{\sc \au{Garc{\'\i}a, FJ} \& \au{Gonz{\'a}lez, H}} \yr{2008}  \at{Normal-mode
  linear analysis and initial conditions of capillary jets}.  \jt{Journal of
  Fluid Mechanics}  \bvol{602},  \pg{81--117}.

\bibitem[Goren(1962)]{goren1962instability}
{\sc \au{Goren, Simon~L}} \yr{1962}  \at{The instability of an annular thread
  of fluid}.  \jt{Journal of Fluid Mechanics}  \bvol{12}~(2),  \pg{309--319}.

\bibitem[Haynes {\em et~al.\/}(2018)Haynes, Vega, Herrada, Benilov \&
  Montanero]{haynes2018stabilization}
{\sc \au{Haynes, M}, \au{Vega, EJ}, \au{Herrada, MA}, \au{Benilov, ES} \&
  \au{Montanero, JM}} \yr{2018}  \at{Stabilization of axisymmetric liquid
  bridges through vibration-induced pressure fields}.  \jt{Journal of colloid
  and interface science}  \bvol{513},  \pg{409--417}.

\bibitem[Holt \& Trinh(1996)]{holt1996faraday}
{\sc \au{Holt, R~Glynn} \& \au{Trinh, Eugene~H}} \yr{1996}  \at{Faraday wave
  turbulence on a spherical liquid shell}.  \jt{Physical review letters}
  \bvol{77}~(7),  \pg{1274}.

\bibitem[Jacqmin \& Duval(1988)]{jacqmin1988instabilities}
{\sc \au{Jacqmin, David} \& \au{Duval, Walter~MB}} \yr{1988}  \at{Instabilities
  caused by oscillating accelerations normal to a viscous fluid-fluid
  interface}.  \jt{Journal of Fluid Mechanics}  \bvol{196},  \pg{495--511}.

\bibitem[Kudrolli \& Gollub(1996)]{kudrolli1996patterns}
{\sc \au{Kudrolli, A} \& \au{Gollub, Jerry~P}} \yr{1996}  \at{Patterns and
  spatiotemporal chaos in parametrically forced surface waves: a systematic
  survey at large aspect ratio}.  \jt{Physica D: Nonlinear Phenomena}
  \bvol{97}~(1-3),  \pg{133--154}.

\bibitem[Kumar \& Tuckerman(1994)]{kumar1994parametric}
{\sc \au{Kumar, Krishna} \& \au{Tuckerman, Laurette~S}} \yr{1994}
  \at{Parametric instability of the interface between two fluids}.  \jt{Journal
  of Fluid Mechanics}  \bvol{279},  \pg{49--68}.

\bibitem[Kumar(2000)]{PhysRevE.62.1416}
{\sc \au{Kumar, Satish}} \yr{2000}  \at{Mechanism for the faraday instability
  in viscous liquids}.  \jt{Phys. Rev. E}  \bvol{62},  \pg{1416--1419}.

\bibitem[Liu \& Liu(2006)]{liu2006linear}
{\sc \au{Liu, Zhihao} \& \au{Liu, Zhengbai}} \yr{2006}  \at{Linear analysis of
  three-dimensional instability of non-newtonian liquid jets}.  \jt{Journal of
  Fluid Mechanics}  \bvol{559},  \pg{451--459}.

\bibitem[Lowry \& Steen(1994)]{lowry1994stabilization}
{\sc \au{Lowry, BJ} \& \au{Steen, PH}} \yr{1994}  \at{Stabilization of an
  axisymmetric liquid bridge by viscous flow}.  \jt{International journal of
  multiphase flow}  \bvol{20}~(2),  \pg{439--443}.

\bibitem[Lowry \& Steen(1995)]{lowry1995flow}
{\sc \au{Lowry, Brian~J} \& \au{Steen, Paul~H}} \yr{1995}  \at{Flow-influenced
  stabilization of liquid columns}.  \jt{Journal of colloid and interface
  science}  \bvol{170}~(1),  \pg{38--43}.

\bibitem[Lowry \& Steen(1997)]{lowry1997stability}
{\sc \au{Lowry, Brian~J} \& \au{Steen, Paul~H}} \yr{1997}  \at{Stability of
  slender liquid bridges subjected to axial flows}.  \jt{Journal of Fluid
  Mechanics}  \bvol{330},  \pg{189--213}.

\bibitem[Maity(2021)]{maity2021floquet}
{\sc \au{Maity, Dilip~Kumar}} \yr{2021}  \at{Floquet analysis on a viscous
  cylindrical fluid surface subject to a time-periodic radial acceleration}.
  \jt{Theoretical and Computational Fluid Dynamics}  \bvol{35}~(1),
  \pg{93--107}.

\bibitem[Maity {\em et~al.\/}(2020)Maity, Kumar \&
  Khastgir]{maity2020instability}
{\sc \au{Maity, Dilip~Kumar}, \au{Kumar, Krishna} \& \au{Khastgir,
  Sugata~Pratik}} \yr{2020}  \at{Instability of a horizontal water
  half-cylinder under vertical vibration}.  \jt{Experiments in Fluids}
  \bvol{61}~(2),  \pg{1--9}.

\bibitem[Marqu{\'e}s(1990)]{marques1990boundary}
{\sc \au{Marqu{\'e}s, Francisco}} \yr{1990}  \at{On boundary conditions for
  velocity potentials in confined flows: Application to couette flow}.
  \jt{Physics of Fluids A: Fluid Dynamics}  \bvol{2}~(5),  \pg{729--737}.

\bibitem[Marr-Lyon {\em et~al.\/}(1997)Marr-Lyon, Thiessen \&
  Marston]{marr1997stabilization}
{\sc \au{Marr-Lyon, Mark~J}, \au{Thiessen, David~B} \& \au{Marston, Philip~L}}
  \yr{1997}  \at{Stabilization of a cylindrical capillary bridge far beyond the
  rayleigh--plateau limit using acoustic radiation pressure and active
  feedback}.  \jt{Journal of Fluid Mechanics}  \bvol{351},  \pg{345--357}.

\bibitem[Marr-Lyon {\em et~al.\/}(2001)Marr-Lyon, Thiessen \&
  Marston]{marr2001passive}
{\sc \au{Marr-Lyon, Mark~J}, \au{Thiessen, David~B} \& \au{Marston, Philip~L}}
  \yr{2001}  \at{Passive stabilization of capillary bridges in air with
  acoustic radiation pressure}.  \jt{Physical review letters}  \bvol{86}~(11),
  \pg{2293}.

\bibitem[Matthiessen(1868)]{matthiessen1868akustische}
{\sc \au{Matthiessen, Ludwig}} \yr{1868}  \at{Akustische versuche, die
  kleinsten transversalwellen der fl{\"u}ssigkeiten betreffend}.  \jt{Annalen
  der Physik}  \bvol{210}~(5),  \pg{107--117}.

\bibitem[Melde(1860)]{melde1860ueber}
{\sc \au{Melde, Franz}} \yr{1860}  \at{Ueber die erregung stehender wellen
  eines fadenf{\"o}rmigen k{\"o}rpers}.  \jt{Annalen der Physik}
  \bvol{187}~(12),  \pg{513--537}.

\bibitem[Moldavsky {\em et~al.\/}(2007)Moldavsky, Fichman \&
  Oron]{moldavsky2007dynamics}
{\sc \au{Moldavsky, Len}, \au{Fichman, Mati} \& \au{Oron, Alexander}} \yr{2007}
   \at{Dynamics of thin liquid films falling on vertical cylindrical surfaces
  subjected to ultrasound forcing}.  \jt{Physical Review E}  \bvol{76}~(4),
  \pg{045301}.

\bibitem[Mollot {\em et~al.\/}(1993)Mollot, Tsamopoulos, Chen \&
  Ashgriz]{mollot1993nonlinear}
{\sc \au{Mollot, DJ}, \au{Tsamopoulos, J}, \au{Chen, T-Y} \& \au{Ashgriz, N}}
  \yr{1993}  \at{Nonlinear dynamics of capillary bridges: experiments}.
  \jt{Journal of fluid mechanics}  \bvol{255},  \pg{411--435}.

\bibitem[Mostert \& Deike(2020)]{mostert_deike_2020}
{\sc \au{Mostert, W.} \& \au{Deike, L.}} \yr{2020}  \at{Inertial energy
  dissipation in shallow-water breaking waves}.  \jt{Journal of Fluid
  Mechanics}  \bvol{890},  \pg{A12}.

\bibitem[Nicol{\'a}s(1992)]{nicolas1992magnetohydrodynamic}
{\sc \au{Nicol{\'a}s, JA}} \yr{1992}  \at{Magnetohydrodynamic stability of
  cylindrical liquid bridges under a uniform axial magnetic field}.
  \jt{Physics of Fluids A: Fluid Dynamics}  \bvol{4}~(11),  \pg{2573--2577}.

\bibitem[Patankar {\em et~al.\/}(2020)Patankar, Basak, , Farsoiya \&
  Dasgupta]{GFM19}
{\sc \au{Patankar, S.}, \au{Basak, S.}, , \au{Farsoiya, P.~K.} \& \au{Dasgupta,
  R.}} \yr{2020} {Viscous stabilisation of Rayleigh-Plateau modes on a
  cylindrical filament through radial oscillatory forcing}.
  \url{https://gfm.aps.org/meetings/dfd-2020/5f5f0e8d199e4c091e67bdbd}, 73TH
  ANNUAL MEETING OF THE APS DIVISION OF FLUID DYNAMICS.

\bibitem[Patankar {\em et~al.\/}(2019)Patankar, Basak \&
  Dasgupta]{patankar2019fragmenting}
{\sc \au{Patankar, Sagar}, \au{Basak, Saswata} \& \au{Dasgupta, Ratul}}
  \yr{2019} Fragmenting a viscous cylindrical fluid filament using the faraday
  instability.  \bt{In {\em APS Division of Fluid Dynamics Meeting
  Abstracts\/}},  \pg{pp. S34--001}.

\bibitem[Patankar {\em et~al.\/}(2018)Patankar, Farsoiya \&
  Dasgupta]{patankar2018faraday}
{\sc \au{Patankar, Sagar}, \au{Farsoiya, Palas~Kumar} \& \au{Dasgupta, Ratul}}
  \yr{2018}  \at{Faraday waves on a cylindrical fluid filament--generalised
  equation and simulations}.  \jt{Journal of Fluid Mechanics}  \bvol{857},
  \pg{80--110}.

\bibitem[Piriz {\em et~al.\/}(2010)Piriz, Prieto, Diaz, Cela \&
  Tahir]{piriz2010dynamic}
{\sc \au{Piriz, AR}, \au{Prieto, G~Rodriguez}, \au{Diaz, I~Mu{\~n}oz},
  \au{Cela, JJ~Lopez} \& \au{Tahir, NA}} \yr{2010}  \at{Dynamic stabilization
  of rayleigh-taylor instability in newtonian fluids}.  \jt{Physical Review E}
  \bvol{82}~(2),  \pg{026317}.

\bibitem[Plateau(1873{\natexlab{{\em a\/}}})]{plateau1873experimental}
{\sc \au{Plateau, Joseph}} \yr{1873{\natexlab{{\em a\/}}}}  \at{Experimental
  and theoretical statics of liquids subject to molecular forces only} .

\bibitem[Plateau(1873{\natexlab{{\em b\/}}})]{plateau1873statique}
{\sc \au{Plateau, Joseph Antoine~Ferdinand}} \yr{1873{\natexlab{{\em b\/}}}}
  {\em Statique exp{\'e}rimentale et th{\'e}orique des liquides soumis aux
  seules forces mol{\'e}culaires\/}, ,  \vol{vol.~2}.  \publ{Gauthier-Villars}.

\bibitem[Popinet(2014)]{popinet2014basilisk}
{\sc \au{Popinet, Stephane}} \yr{2014} Basilisk. \url{http://basilisk.fr}.

\bibitem[Prosperetti(1976)]{prosperetti1976viscous}
{\sc \au{Prosperetti, Andrea}} \yr{1976}  \at{Viscous effects on
  small-amplitude surface waves}.  \jt{Physics of Fluids (1958-1988)}
  \bvol{19}~(2),  \pg{195--203}.

\bibitem[Prosperetti(2011)]{prosperetti2011advanced}
{\sc \au{Prosperetti, Andrea}} \yr{2011} {\em Advanced mathematics for
  applications\/}.  \publ{Cambridge University Press}.

\bibitem[Raco(1968)]{raco1968electrically}
{\sc \au{Raco, Roland~J}} \yr{1968}  \at{Electrically supported column of
  liquid}.  \jt{Science}  \bvol{160}~(3825),  \pg{311--312}.

\bibitem[Raman(1909)]{raman1909maintenance}
{\sc \au{Raman, CV}} \yr{1909}  \at{The maintenance of forced oscillations of a
  new type}.  \jt{Nature}  \bvol{82}~(2093),  \pg{156--157}.

\bibitem[Raman(1912)]{raman1912experimental}
{\sc \au{Raman, Chandrasekhara~Venkata}} \yr{1912}  \at{Experimental
  investigations on the maintenance of vibrations}.  \jt{Proc. Indian
  Association for the Cultivation of Sci. Bulletin 6} .

\bibitem[Rayleigh(1878)]{rayleigh1878instability}
{\sc \au{Rayleigh, Lord}} \yr{1878}  \at{On the instability of jets}.
  \jt{Proceedings of the London mathematical society}  \bvol{1}~(1),
  \pg{4--13}.

\bibitem[Rayleigh(1883)]{rayleigh1883xxxiii}
{\sc \au{Rayleigh, Lord}} \yr{1883}  \at{Xxxiii. on maintained vibrations}.
  \jt{The London, Edinburgh, and Dublin Philosophical Magazine and Journal of
  Science}  \bvol{15}~(94),  \pg{229--235}.

\bibitem[Rayleigh(1887)]{rayleigh1887xvii}
{\sc \au{Rayleigh, Lord}} \yr{1887}  \at{Xvii. on the maintenance of vibrations
  by forces of double frequency, and on the propagation of waves through a
  medium endowed with a periodic structure}.  \jt{The London, Edinburgh, and
  Dublin Philosophical Magazine and Journal of Science}  \bvol{24}~(147),
  \pg{145--159}.

\bibitem[Rayleigh(1892{\natexlab{{\em a\/}}})]{rayleigh1892instability}
{\sc \au{Rayleigh, L}} \yr{1892{\natexlab{{\em a\/}}}}  \at{On the instability
  of a cylinder of viscous liquid under capillary force. philosophical
  magazine} .

\bibitem[Rayleigh(1892{\natexlab{{\em b\/}}})]{rayleigh1892xvi}
{\sc \au{Rayleigh, Lord}} \yr{1892{\natexlab{{\em b\/}}}}  \at{Xvi. on the
  instability of a cylinder of viscous liquid under capillary force}.  \jt{The
  London, Edinburgh, and Dublin Philosophical Magazine and Journal of Science}
  \bvol{34}~(207),  \pg{145--154}.

\bibitem[Rohlfs {\em et~al.\/}(2014)Rohlfs, Binz \&
  Kneer]{rohlfs2014stabilizing}
{\sc \au{Rohlfs, Wilko}, \au{Binz, Matthias} \& \au{Kneer, Reinhold}} \yr{2014}
   \at{On the stabilizing effect of a liquid film on a cylindrical core by
  oscillatory motions}.  \jt{Physics of Fluids}  \bvol{26}~(2),  \pg{022101}.

\bibitem[Rutland \& Jameson(1971)]{rutland1971non}
{\sc \au{Rutland, DF} \& \au{Jameson, GJ}} \yr{1971}  \at{A non-linear effect
  in the capillary instability of liquid jets}.  \jt{Journal of Fluid
  Mechanics}  \bvol{46}~(2),  \pg{267--271}.

\bibitem[Sankaran \& Saville(1993)]{sankaran1993experiments}
{\sc \au{Sankaran, Subramanian} \& \au{Saville, DA}} \yr{1993}  \at{Experiments
  on the stability of a liquid bridge in an axial electric field}.  \jt{Physics
  of Fluids A: Fluid Dynamics}  \bvol{5}~(4),  \pg{1081--1083}.

\bibitem[Sanz(1985)]{sanz_1985}
{\sc \au{Sanz, Angel}} \yr{1985}  \at{The influence of the outer bath in the
  dynamics of axisymmetric liquid bridges}.  \jt{Journal of Fluid Mechanics}
  \bvol{156},  \pg{101–140}.

\bibitem[Shats {\em et~al.\/}(2014)Shats, Francois, Xia \&
  Punzmann]{shats2014turbulence}
{\sc \au{Shats, Michael}, \au{Francois, Nicolas}, \au{Xia, Hua} \&
  \au{Punzmann, Horst}} \yr{2014} Turbulence driven by faraday surface waves.
  \bt{In {\em International Journal of Modern Physics: Conference Series\/}}, ,
   \vol{vol.~34},  \pg{p. 1460379}. World Scientific.

\bibitem[Singh {\em et~al.\/}(2019)Singh, Farsoiya \& Dasgupta]{singh2019test}
{\sc \au{Singh, Manpreet}, \au{Farsoiya, Palas~Kumar} \& \au{Dasgupta, Ratul}}
  \yr{2019}  \at{Test cases for comparison of two interfacial solvers}.
  \jt{International Journal of Multiphase Flow} .

\bibitem[Song {\em et~al.\/}(2020)Song, Kartawira, Hillaire, Li, Eaker, Kiani,
  Daniels \& Dickey]{song2020overcoming}
{\sc \au{Song, Minyung}, \au{Kartawira, Karin}, \au{Hillaire, Keith~D}, \au{Li,
  Cheng}, \au{Eaker, Collin~B}, \au{Kiani, Abolfazl}, \au{Daniels, Karen~E} \&
  \au{Dickey, Michael~D}} \yr{2020}  \at{Overcoming rayleigh--plateau
  instabilities: Stabilizing and destabilizing liquid-metal streams via
  electrochemical oxidation}.  \jt{Proceedings of the National Academy of
  Sciences}  \bvol{117}~(32),  \pg{19026--19032}.

\bibitem[Sterman-Cohen {\em et~al.\/}(2017)Sterman-Cohen, Bestehorn \&
  Oron]{sterman2017rayleigh}
{\sc \au{Sterman-Cohen, Elad}, \au{Bestehorn, Michael} \& \au{Oron, Alexander}}
  \yr{2017}  \at{Rayleigh-taylor instability in thin liquid films subjected to
  harmonic vibration}.  \jt{Physics of Fluids}  \bvol{29}~(5),  \pg{052105}.

\bibitem[Stone {\em et~al.\/}(2004)Stone, Stroock \&
  Ajdari]{stone2004engineering}
{\sc \au{Stone, Howard~A}, \au{Stroock, Abraham~D} \& \au{Ajdari, Armand}}
  \yr{2004}  \at{Engineering flows in small devices: microfluidics toward a
  lab-on-a-chip}.  \jt{Annu. Rev. Fluid Mech.}  \bvol{36},  \pg{381--411}.

\bibitem[Taylor(1969)]{taylor1969electrically}
{\sc \au{Taylor, Geoffrey~Ingram}} \yr{1969}  \at{Electrically driven jets}.
  \jt{Proceedings of the Royal Society of London. A. Mathematical and Physical
  Sciences}  \bvol{313}~(1515),  \pg{453--475}.

\bibitem[Thiele {\em et~al.\/}(2006)Thiele, Vega \& Knobloch]{thiele2006long}
{\sc \au{Thiele, Uwe}, \au{Vega, Jose~M} \& \au{Knobloch, Edgar}} \yr{2006}
  \at{Long-wave marangoni instability with vibration}.  \jt{Journal of Fluid
  Mechanics}  \bvol{546},  \pg{61--87}.

\bibitem[Thiessen {\em et~al.\/}(2002)Thiessen, Marr-Lyon \&
  Marston]{thiessen2002active}
{\sc \au{Thiessen, David~B}, \au{Marr-Lyon, Mark~J} \& \au{Marston, Philip~L}}
  \yr{2002}  \at{Active electrostatic stabilization of liquid bridges in low
  gravity}.  \jt{Journal of Fluid Mechanics}  \bvol{457},  \pg{285--294}.

\bibitem[Troyon \& Gruber(1971)]{troyon1971theory}
{\sc \au{Troyon, Francis} \& \au{Gruber, Ralf}} \yr{1971}  \at{Theory of the
  dynamic stabilization of the rayleigh-taylor instability}.  \jt{The Physics
  of Fluids}  \bvol{14}~(10),  \pg{2069--2073}.

\bibitem[Tyndall(1901)]{tyndall1901sound}
{\sc \au{Tyndall, John}} \yr{1901} {\em Sound\/}, ,  \vol{vol.~7}.
  \publ{Collier}.

\bibitem[Vega \& Montanero(2009)]{vega2009damping}
{\sc \au{Vega, EJ} \& \au{Montanero, JM}} \yr{2009}  \at{Damping of linear
  oscillations in axisymmetric liquid bridges}.  \jt{Physics of Fluids}
  \bvol{21}~(9),  \pg{092101}.

\bibitem[Wang {\em et~al.\/}(2005)Wang, Joseph \& Funada]{wang2005pressure}
{\sc \au{Wang, Jing}, \au{Joseph, Daniel~D} \& \au{Funada, Toshio}} \yr{2005}
  \at{Pressure corrections for potential flow analysis of capillary instability
  of viscous fluids}.  \jt{Journal of Fluid Mechanics}  \bvol{522},
  \pg{383--394}.

\bibitem[Weber(1931)]{weber1931zerfall}
{\sc \au{Weber, Constantin}} \yr{1931}  \at{Zum zerfall eines
  fl{\"u}ssigkeitsstrahles}.  \jt{ZAMM-Journal of Applied Mathematics and
  Mechanics/Zeitschrift f{\"u}r Angewandte Mathematik und Mechanik}
  \bvol{11}~(2),  \pg{136--154}.

\bibitem[Wolf(1970)]{wolf1970dynamic}
{\sc \au{Wolf, GH}} \yr{1970}  \at{Dynamic stabilization of the interchange
  instability of a liquid-gas interface}.  \jt{Physical Review Letters}
  \bvol{24}~(9),  \pg{444}.

\bibitem[Wolf(1969)]{wolf1969dynamic}
{\sc \au{Wolf, Gerhard~Hans}} \yr{1969}  \at{The dynamic stabilization of the
  rayleigh-taylor instability and the corresponding dynamic equilibrium}.
  \jt{Zeitschrift f{\"u}r Physik A Hadrons and nuclei}  \bvol{227}~(3),
  \pg{291--300}.

\bibitem[{Wolfram Research, Inc.}(2017)]{mathematica2017}
{\sc \au{{Wolfram Research, Inc.}}} \yr{2017} Mathematica 11.

\bibitem[Woods \& Lin(1995)]{woods1995instability}
{\sc \au{Woods, David~R} \& \au{Lin, SP}} \yr{1995}  \at{Instability of a
  liquid film flow over a vibrating inclined plane}.  \jt{Journal of Fluid
  Mechanics}  \bvol{294},  \pg{391--407}.

\bibitem[Yih(1967)]{yih1967instability}
{\sc \au{Yih, Chia-Shun}} \yr{1967}  \at{Instability due to viscosity
  stratification}.  \jt{Journal of Fluid Mechanics}  \bvol{27}~(2),
  \pg{337--352}.

\end{thebibliography}

\end{document}